\documentclass[journal=jpclcd,manuscript=communication]{achemso}

\usepackage{multicol}
\usepackage{lipsum} 
\usepackage[]{mhchem} 

\usepackage{color}
\usepackage{pdfpages}

\author{Swen Helstroffer}
\affiliation{Université de Strasbourg, CNRS, Institut Charles Sadron UPR22, F-67000 Strasbourg, France}

\author{Ludovic Gardr\'e}
\affiliation{Universit\'e Claude Bernard Lyon 1, CNRS, Institut Lumi\`ere Mati\`ere, UMR5306, F69100 Villeurbanne, France}

\author{Giovanna Fragneto}
\affiliation{Institut Laue-Langevin, 71 av. des Martyrs, BP 156, 38042 Grenoble Cedex, France}
\affiliation{current address: European Spallation source ERIC, Partikelgatan 2, P.O. Box 176, SE-221 00 Lund, Sweden}

\author{Arnaud Hemmerle}
\affiliation{Synchrotron SOLEIL, L'Orme des Merisiers, Départementale 128, 91190 Saint-Aubin, France}

\author{Léo Henry}
\affiliation{Université de Strasbourg, CNRS, Institut Charles Sadron UPR22, F-67000 Strasbourg, France}

\author{Laurent Joly}
\affiliation{Universit\'e Claude Bernard Lyon 1, CNRS, Institut Lumi\`ere Mati\`ere, UMR5306, F69100 Villeurbanne, France}

\author{Fabrice Thalmann}
\affiliation{Université de Strasbourg, CNRS, Institut Charles Sadron UPR22, F-67000 Strasbourg, France}

\author{Claire Loison}
\affiliation{Universit\'e Claude Bernard Lyon 1, CNRS, Institut Lumi\`ere Mati\`ere, UMR5306, F69100 Villeurbanne, France}

\author{Pierre Muller}
\affiliation{Université de Strasbourg, CNRS, Institut Charles Sadron UPR22, F-67000 Strasbourg, France}

\author{Thierry Charitat}
\affiliation{Université de Strasbourg, CNRS, Institut Charles Sadron UPR22, F-67000 Strasbourg, France}
\email{charitat@unistra.fr}

\title[An \textsf{achemso} demo]
  {The role of confined water in the emergence of electrostatic strong coupling as revealed by nanoseparated charged lipid layers}

\abbreviations{IR,NMR,UV}

\begin{document}

\begin{tocentry}






\includegraphics[width=5.15cm]{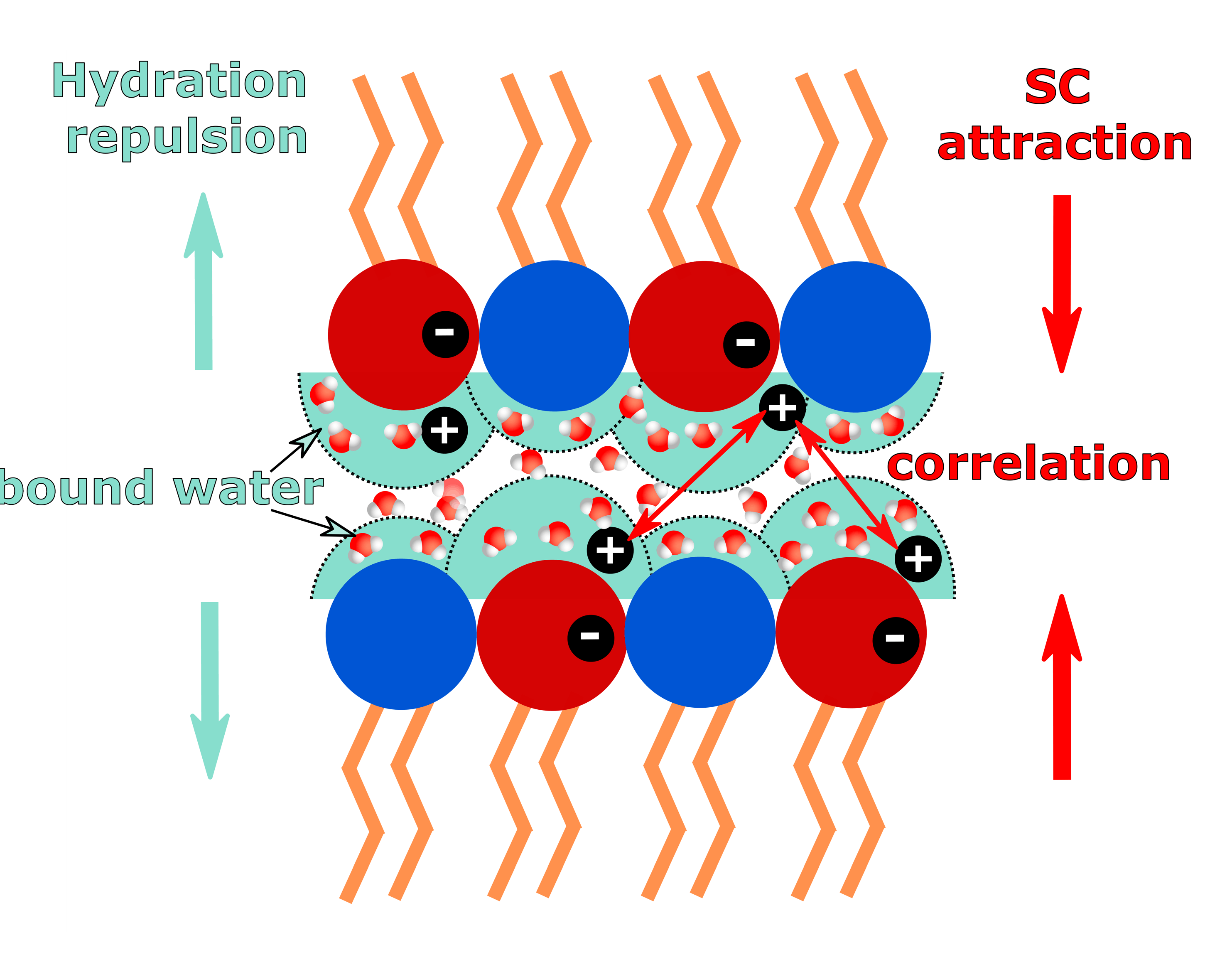}

\end{tocentry}


\begin{abstract}

This study investigates the interplay between Strong Coupling (SC) attraction and hydration repulsion in nanoconfined water between like-charged phospholipid layers. We combine reflectivities with numerical simulations to analyze supported phospholipid layers under different relative humidity and surface charge densities. X-ray fluorescence demonstrates that we can control the valence of the associated counterions. Experimental measurement of the water thickness, combined with precise determination of charged surface positions by numerical simulations, clearly demonstrates the existence of electrostatic attraction in the presence of monovalent counterions. It enables us to compare our experiments with a theoretical model showing that charge-screening by hydration water induces SC attraction, even at moderate surface charge densities with monovalent counterions. Furthermore, hydration repulsion is stronger for DPPS compared to DPPC. These findings offer insights into the forces that control interactions between phospholipid layers and have important implications for biological and colloidal systems.

\end{abstract}

\section{Introduction}

Surfaces immersed in water naturally acquire charges, leading to complex interactions mediated by mobile ions. The classical approach to address this many-body problem is through a mean-field description, known as the Poisson-Boltzmann or Weak Coupling (WC) theory \cite{debye1923theorie,herrero2024poisson}. A key feature of the WC theory is that it predicts exclusively repulsive interactions between like-charged bodies \cite{ANDELMAN1995603}.
However, when ionic correlation becomes significant, the WC theory breaks down and the system may enter the Strong Coupling (SC) regime, where correlations can lead to electrostatic attraction between like-charged surfaces\cite{KJELLANDER198449,rouzina1996,Moreira2000,Netz2001Aug,NAJI2005131,Naji2013Oct,samaj2018}. Although counterintuitive, this attraction appears to be quite universal and is necessary to explain the cohesion of cement\cite{Goyal2021Aug}, biopolymers such as DNA\cite{Bloomfield1991Nov} or microtubules\cite{Wong2006Dec}, mica surfaces\cite{Kekicheff1993Oct}, vesicles\cite{KOMOROWSKI2018,Komorowski2020May}, and layers of charged phospholipids\cite{Mukhina2019Nov}. Until recently, the SC regime was observed only for systems with multivalent counterions, but recent studies have demonstrated that in nanoconfined water, monovalent counterions can also induce attraction \cite{Mukhina2019Nov,Goyal2021Aug,Palaia2022Apr}. This occurs as the out-of-plane dielectric constant of the confined water approaches its value at optical frequencies (optical limit) \cite{Fumagalli2018Jun,Michaelides2024}, thus enhancing the ionic correlations. This study focuses on exploring the conditions under which SC attraction appears in such nanoconfined environments.

\begin{figure}[h!]
\centering
\includegraphics[width=0.45\textwidth]{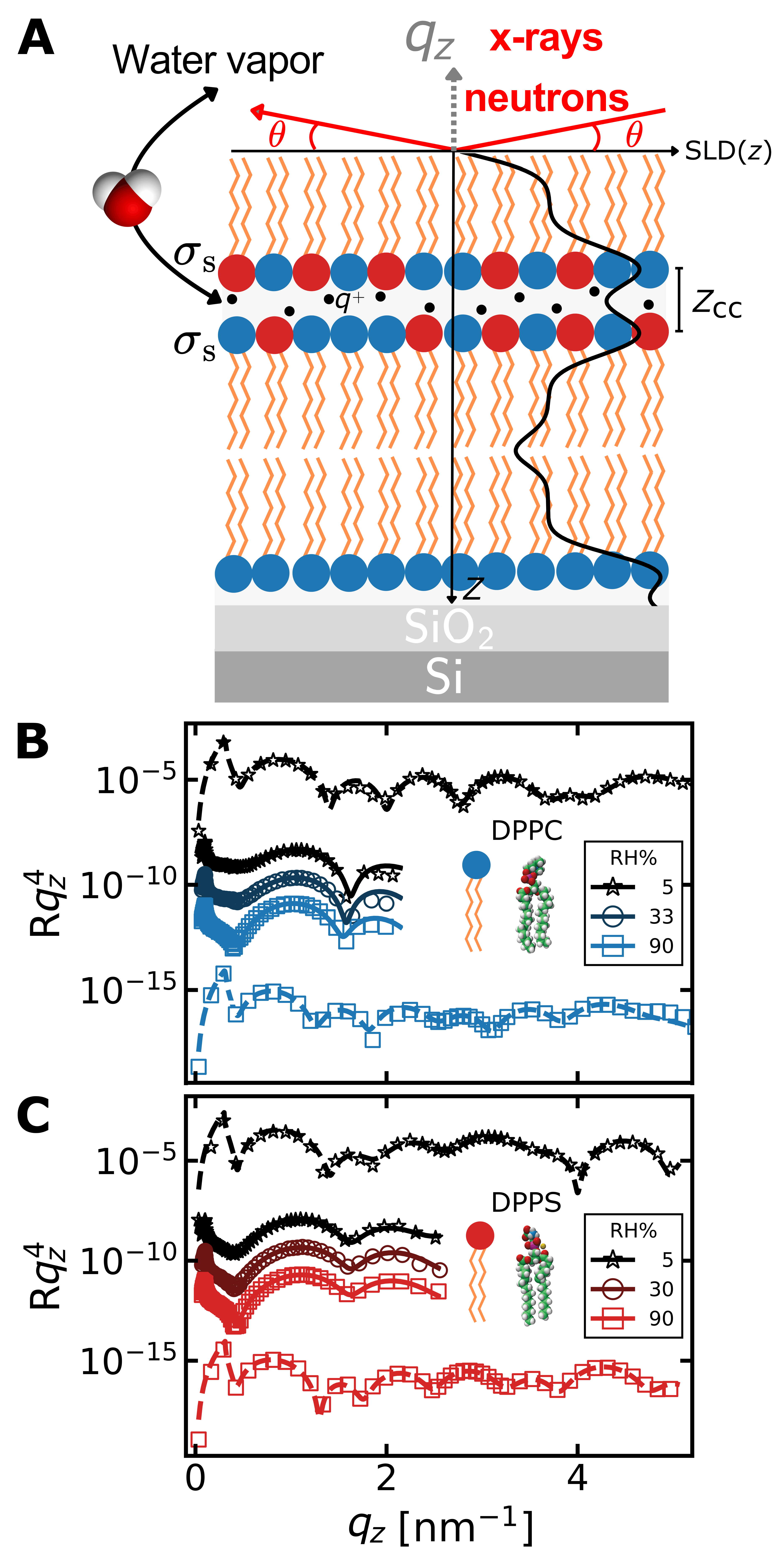}
\caption{(A) Schematic representation of a charged trilayer with a mixture of DPPS (red head) and DPPC (blue head) and a corresponding negative surface charge density, $\sigma_{\rm{S}}$. The DPPS heads are associated with counter ions (black dots). The incoming and reflected specular beams are shown in red, along with the transfer wave vector $q_z = 4\pi/\lambda \sin \theta$. The distance between the charged planes $z_{\rm cc}$ is shown schematically. (B and C) Neutron (NR) and X-Ray Reflectivity (XRR) data are presented as $Rq_z^4$ on a semilog scale. NR data show $\rm{D_2O}$ contrast, while XRR is performed with $\rm{H_2O}$. (B) DPPC and (C) DPPS at RH ranging from 5\% to 90\% (color gradient). Data are shifted by 1.5 decades for clarity. Lines represent the best fits corresponding to the scattering length density profiles in Figure \ref{fig:2}.}
\label{fig:1}
\end{figure}

Zwitterionic phospholipids in the multilamellar phase have long been a valuable model for physicists to probe neutral surface interactions through nanoconfined water\cite{Lis1982Mar}. The flat geometry of this model system and the ability to perform pressure-distance measurements make it ideal for comparison with theoretical models. It is particularly adapted for studying the hydration force, a repulsive force that appears between polar surfaces\cite{Israelachvili2011}. This repulsion is known to prevent the collapse of hydrated matter via the dominating van der Waals attraction at separations below 2 nm. Initially thought to be universal and dependent only on the structuring of water near interfaces, it has recently been shown to be surface-specific\cite{Schlaich2024Apr}. Charged phospholipids exhibit enhanced water ordering around their heads\cite{Schlaich2019Jan,Chen2010Aug}, but to our knowledge, there has been no quantitative analysis of their hydration repulsion. Therefore, our motivations in this paper are threefold: we aim to (1) quantify hydration repulsion for negatively charged phospholipids, (2) determine how this force balances with SC attraction, and (3) decipher when SC attraction appears in phospholipid layer interactions in presence of monovalent or divalent counterions. 

Our strategy was to study mixtures of negatively charged (DPPS) and zwitterionic (DPPC) phospholipids at different mixing ratio, in order to continuously tune the surface charge, 
deposited on the silica surface of a silicon single crystal substrate. We tuned the osmotic pressure in the water layers by controlling the relative humidity (RH), which allowed us to reach high negative pressures that drive the water molecules out of the lipid layers. By analyzing neutron (NR) and X-ray (XRR) reflectivities, we extracted pressure-distance curves for different charge fractions and relative humidity. We were able to completely describe the behavior of our systems using a model that accounts for the balance between lipid hydration repulsion and SC attraction.

\section{Materials and Methods}

\subsection{Samples preparation}

The samples consisted of an odd number of phospholipid layers deposited at the air/silicon interface. The substrates were polished to be atomically smooth, as described previously \cite{Hemmerle2012Dec}. We used zwitterionic phospholipid DPPC (1,2-dipalmitoyl-sn-glycero-3-phosphocholine, Avanti Polar Lipids, Alabaster, AL; main transition temperature $T_{\rm{m}} = 41\, ^\circ \text{C}$) and anionic phospholipid DPPS (1,2-dipalmitoyl-sn-glycero-3-phospho-L-serine (sodium salt), Avanti Polar Lipids, Alabaster, AL; main transition temperature $T_{\rm{m}} = 54\, ^\circ \text{C}$).
The first layer transferred to the substrate was always composed solely of a pure PC, since zwitterionic phospholipids adhere more strongly to the substrate \cite{Mukhina2019Nov}, improving the quality of deposition. For the two next layers, we studied eight different DPPC/DPPS mixtures with varying DPPS molar fractions \( j = n_{\rm{DPPS}} / (n_{\rm{DPPS}} + n_{\rm{DPPC}}) \), with \( j = 0, 0.2, 0.4, 0.5, 0.6, 0.7, 0.8\), and 1. All samples were prepared using the Langmuir-Blodgett (LB) technique. For neutral phospholipids, LB generally limits the deposition to three layers (trilayer)\cite{Charitat1999}. However, for DPPS fractions greater than 0.6, we were able to stack more than three layers, indicative of the SC attraction regime. We limited ourselves to five layers (pentalayer) at $j=0.7$ and $j=0.8$. Further details of the experimental procedure are provided in the Supporting Information (Section 1).

\subsection{Neutrons and x-rays reflectivities}

We used neutron reflectivity and X-ray reflectivity to achieve high-resolution characterization ($\sim 0.1$ nm) of the trilayer and pentalayer structures. NR measurements were conducted on the D17 reflectometer at the Institut Laue-Langevin (ILL) in Grenoble, France (wavelength $\lambda = 0.2-2$ nm) \cite{cubitt2002}, and XRR experiments were performed on the SIRIUS beamline \cite{hemmerle2024SIRIUS} at Synchrotron SOLEIL in Paris-Saclay, France, using an 8 keV X-ray beam (wavelength $\lambda = 0.155$ nm). A scheme of the reflectivity setup is shown in Figure \ref{fig:1}A.
The reflectivity $R(q_z)$ is defined as the ratio of the intensities of the specular reflected beam to the incoming beam. $R(q_z)$ is expressed as a function of the wave vector transfer, $q_z = 4\pi\sin{\left(\theta\right)}/\lambda$, where $\theta$ is the grazing angle of incidence and reflection.  NR and XRR data were fitted using the {\it refnx} python package \cite{Nelson2019Feb}, employing a slab model where interfaces are represented by error functions. This method allows the construction of the scattering length density (SLD) profile, which is then related to the reflectivity $R(q_z)$ using the Abeles method \cite{heavens1960optical}. The fitting routine employs least-squares minimization and Bayesian Markov-chain Monte Carlo sampling. Further details on the fitting procedure and model constraints are provided in the Supporting Information (Section 2). XRR and NR experiments were carried out in humidity chambers with precise RH control\cite{Sebastiani2012}. For NR, we employed two water contrasts (H$_2$O with an SLD of $-0.56 \times 10^{-4}$ nm$^{-2}$ and D$_2$O with an SLD of $6.36 \times 10^{-4}$ nm$^{-2}$), allowing for co-refinement during fitting, which reduced parameter ambiguity.

\subsection{Numerical simulations}

In addition to these experimental approaches, molecular dynamics (MD) simulations were performed using a classical atomistic model that describes the molecular structures, plus electrostatic and van der Waals interactions (CHARMM36 force field\cite{Klauda2010}, more details are provided in Supporting Information, Section 3). The simulation box contains planar lipid layers of pure DPPS in the gel phase (293 K), with their sodium counterions in a confined water layer. It represents the region of the second and third monolayer of the experimental samples. Because of the 3D periodic boundary conditions used in the simulations, the simulated systems are in fact equivalent to stacks of bilayers with varying hydration number HN (number of water molecules per lipid head), from 0 to 45 water molecules per DPPS molecule. For each hydration level, a 1000 ns-equilibration was performed before the 600 ns-production. Two analyses were performed, and the results were compared to experimental data. 

First, SLD profiles in $z$-direction, comparable to the experimental fits, were obtained by summing the contributions of all atoms, calculated by the product of the atomic local densities, and the relevant atomic scattering length. In the case of neutron SLD profiles, calculated in D2O, we had to take into account that the exchange of water and D2O molecules was not perfect.

\begin{figure}[h!]
\centering
\includegraphics[width=0.5\textwidth]{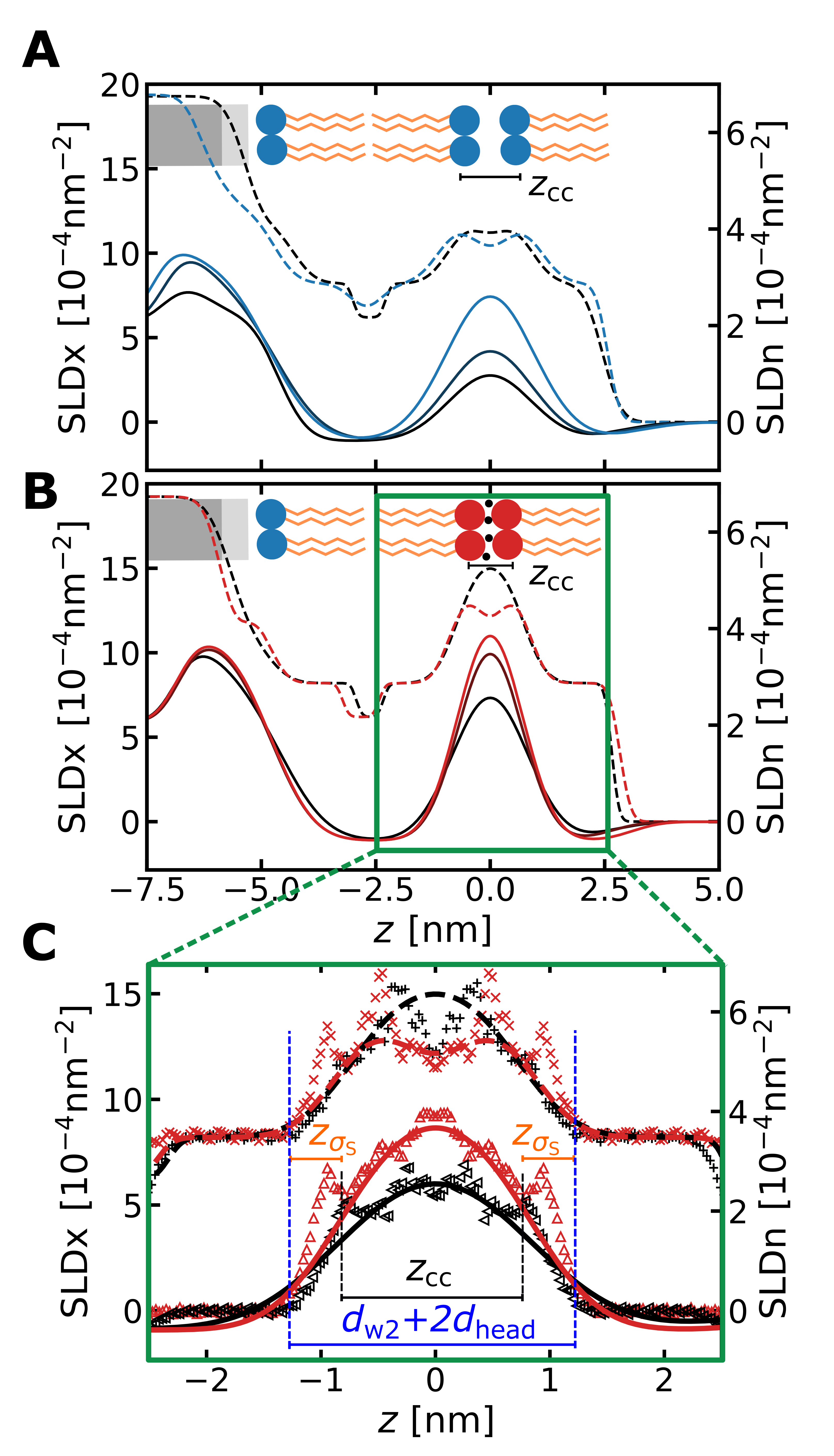}
\caption{Scattering length density (SLD) profiles for NR (solid lines, right axis) and XRR (dashed lines, left axis) for DPPC (A) and DPPS (B) trilayers. The color code follows the one of Figure \ref{fig:1}. (C) Comparison between experimental SLD profiles for NR (solid lines, right axis) and XRR (dashed lines, left axis) zoomed on the second water layer (green square) for RH=5\% (black curve) and RH=90\% (red curve) and SLD profiles from numerical simulations for neutrons (hydration number HN=3 (\textcolor{black}{$\triangleleft$}) and HN=6 (\textcolor{red}{$\triangle$})) and for X-rays (HN=3 ($+$) and HN=6 (\textcolor{red}{$\times$})). All profiles have been centered on the midpoint of the second water layer.}
\label{fig:2}
\end{figure}

Second, water ordering within the confined layer was characterized using polarization profiles, defined as the average projection of water dipole moments along the $z$-axis. Both neutrons and X-ray SLD profiles in $z$-direction were obtained by summing the contributions of all atoms, calculated by the product of the atomic local density, and the relevant atomic scattering length (see Fig. \ref{fig:2}C).

\section{Results}

To interpret the experimental data, we propose to use in the discussion a simple model where two parallel planes of surface charge density $\sigma_{\rm S}$ are separated by a distance $z_{\rm cc}$ in the presence of counterions. In what follows, we justify this approach by discussing the nature of counterions (and in particular their valence) and showing how to determine the two important parameters $\sigma_{\rm S}$ and $z_{\rm cc}$ in our experimental systems.

\subsection{Nature of the counterions and surface charge density}

The issues of charge density and the nature of counterions are subtle. For example, charge density can be influenced by the pH of the solution. The nature of the counterions is influenced by the species present during Langmuir-Blodgett deposition, even in trace amounts. To clarify these two points, we therefore carried out quantitative X-ray fluorescence experiments by adding Cs$^+$ or Ca$^{2+}$ solutions into the subphase (for more details see the Supporting Information, Section 4). These experiments show that all the DPPS heads are charged and neutralized by added cations, as the ratio of cesium ions to phosphorus atoms in DPPS is one, whereas the ratio of calcium ions to phosphate atoms is one half, regardless of the DPPC/DPPS molar ratio. This allows us to determine the surface charge density of DPPS precisely as one electron per PS molecule. X-ray reflectivity also showed no thickness differences between the samples prepared in the presence or absence of Cs$^+$ or with pure water.

In conclusion, we can rule out any significant pollution by multivalent ions. We also precisely know the surface charge density of DPPS lipid layers $\sigma_{\rm S,PS} = 1e^-/ 0.42$ nm$^2$. We can also calculate the total surface charge density of a layer of DPPS fraction $j$ as $\sigma_{\text{S}}\left(j\right) = j\sigma_{\rm S,PS} + (1-j)\sigma_{\rm S,PC}$ with a negligible contribution from DPPC molecules ($\sigma_{\rm S,PC} \sim 10^{-3} \, e^-/\text{nm}^2$) \cite{Pincet1999,Hemmerle2012Dec}.

\subsection{Structure of the trilayer}

Figures \ref{fig:1}B and C present $R(q_z)$ profiles for DPPS molar fractions $j=0$ and $j=1$. We achieved highly accurate fits across all DPPS fractions for both XRR and NR data. For clarity, we only display the D$_2$O reflectivity profiles for these two fractions along with the corresponding SLD profiles in Figure \ref{fig:2}. The remaining reflectivity and SLD profiles are provided in the Supporting Information (Section 2). Noticeably, the sections of the SLD profiles corresponding to the regions modeled by the atomistic simulations closely match the MD predictions (see Figure \ref{fig:2}C). This cross-validation between the two different approaches provides a correspondence between the water hydration number used in the simulations and the experimental water thickness $d_{\rm w2}$. 
Experimentally observed HN, calculated with $n_{\rm w}=A_{\rm l}d_{\rm w2}/2v_{\rm w}$ where $v_{\rm w} = 30$ \AA$^3$ is the molecular volume of water and $A_{\rm l} = 42$ \AA$^2$ the area per lipid, typically range from $3\pm 1$ to $7\pm 1$. These values are comparable to or even lower than the approximately 7 water molecules strongly oriented by a single phosphatidylserine molecule, resulting in a significant reduction of their orientational degrees of freedom~\cite{Miller1999}. We also checked that the area per lipid $A_{\rm l}$ and the head thickness $d_{\rm head}$ remained constant whatever the relative humidity (see Supplementary information, section 2).

The thickness of the first layer of water, $d_{\rm{w1}}$, located between the substrate and the bilayer, remains very small ($< 1 $ nm) and constant over the entire relative humidity range studied. 
We therefore concentrate our study on the second layer of water, located between the two charged lipid layers. X-rays and neutrons interact differently with matter, and the most robust distance for comparing all the experiments is $d_{\rm w2}+2d_{\rm head}$, the distance between the two head-chain interfaces (ester group positions), corresponding to the region where neutron and x-ray SLDs vary the most (see Figure \ref{fig:2}B). We therefore extract $d_{\rm{w2}}+2d_{\rm{head}}$ in Figure \ref{fig:2}A from the SLD profiles, with an RH ranging from very dry conditions (RH = 5\%) to high humidity (RH = 90\%).

\subsection{Distance between charged surfaces $z_{\rm cc}$}

To compare our experimental results with theoretical models, we need to define a distance between the two lipid layers. In the case of charged layers, theoretical descriptions generally use the distance between two virtual charged surfaces located in the lipid heads, which we will denote by $z_{\rm cc}$ (see Fig. \ref{fig:2}C). To calculate $z_{\rm cc}$ using
\begin{equation}
z_{\rm cc}= d_{\rm{w2}}+2\left(d_{\rm{head}}-z_{\rm \sigma_{\rm S}}\right),
\end{equation}
we need to locate the position $z_{\rm \sigma_{\rm S}}$ of the charges measured from the position of the ester groups (see Figure \ref{fig:2}). 

\subsubsection{Charges distribution-Determination of $z_{\rm \sigma_{\rm S}}$}

The charge distribution in charged DPPS head is complex. The phosphate (PO$_4^-$) and carboxyl (CO$_2^-$) groups are negatively charged (charge $-e$) and the ammonium group (NH$_4^+$) is positively charged $+e$. The head as a whole carries a net negative charge $-e$. Depending on the lipid head orientation,  these charges lie more or less deep along the $z$-direction. Numerical simulations presented in Supplementary Information (section 5) show that the barycenters of the positive and negative charges are almost coincident and lie at a distance $z_{\sigma_{\rm S}}\simeq 0.5$ nm from the ester group\footnote{We have chosen to measure these positions relative to the position of the ester group, which marks the transition between the chains and the lipid heads, and corresponds to the zone of maximum change in the contrast (both for SLD$_{\rm X}$ and SLD$_{\rm n}$).}. At first order, It therefore seems reasonable to represent the charge distribution of the heads by a charged plane without thickness, with a net charge of $-e$ per lipid, located at the distance $z_{\sigma_{\rm S}}=0.5$ nm from the ester group, and to neglect the dipolar contribution. For sake of simplicity, we have generalized this definition to the case of neutral layers (j = 0). This means fixing the average position of the lipid heads when calculating the hydration pressure. It only results in a shift in the value of the reference hydration pressure, $P_{\rm h,PC}$.

\subsubsection{Experimental results for $z_{\rm cc}$}

Experimental results for $z_{\rm cc}$ for the different surface charges $j$ and as a function of RH are shown in Figures \ref{fig:3}B and C. At low humidity, $z_{\rm cc}$ reaches a minimum of around \mbox{0.4-0.6 nm}, corresponding to zero water thickness ($z_{\rm cc}\sim 2\left(d_{\rm head}-z_{\sigma_{\rm S}}\right)$). Under these conditions of high osmotic pressure, the phospholipid heads are in close contact, leaving only the water molecules of hydration. As RH increases, osmotic pressure decreases, and $z_{\rm cc}$ increases accordingly. This trend is also clear in the SLD$_{\rm{X}}$ profiles shown in Figure \ref{fig:2}. Finally, Figure \ref{fig:3}B shows that $z_{\rm cc}$ is generally lower in the presence of charged lipids ($j\neq 0$) and Figure \ref{fig:3}C that the variation of $z_{\rm cc}$ with charge fraction is not monotonic, but passes through a maximum for $j\simeq 0.2-0.3$. In the following, we show that one can interpret these results using a simple continuous model describing the interactions between the 2 lipid monolayers.

\section{Discussion}

\subsection{Interactions between lipid layers}

The interactions between the lipid layers that control the water thickness $d_{\rm w2}$ (and therefore $z_{\rm cc}$) are complex. There are direct interactions (head-head, chain-chain, etc.) as well as indirect interactions mediated by solvent molecules or counterions \cite{schneckPNAS(2012)}. In all cases, entropic and enthalpic contributions must be taken into account. Finally, these interactions are also influenced by thermal fluctuations of the membrane. We propose below a minimal model to describe all our experimental results, based on a continuous medium approach describing the interaction energy per unit area $U\left(z\right)$ between lipid layers associated to the disjoining pressure $P=-\partial U/\partial z$. In our configuration, at solid-air interface, the water layers confined between the phospholipids are also in equilibrium with their vapor phase, resulting in a large negative osmotic pressure\cite{Allemand2023Jun},

\begin{equation}
P_{\rm O} = \frac{k_{\rm{B}} T}{v_{\rm w}}\log{\left(\frac{\rm{RH}}{100}\right)},
\label{P_O}
\end{equation}
where $k_{\rm{B}}$ is the Boltzmann constant and $T$ is the temperature. This osmotic pressure adds to the disjoining pressure of the water films and the equilibrium distance between the lipid layers $z_{\rm cc}$ is determined by the balance of the pressures 
\begin{equation}
P_{\rm{O}}+P\left(z_{\rm cc}\right)=0.
\label{eq_pressure_total_0}
\end{equation}
Our systems at the solid-air interface are highly confined by the osmotic pressure, and therefore quite different from studies on fully hydrated bilayers. To go further and describe our experimental results we have made the following highly simplifying assumptions: (i) the various contributions to disjoining pressure are additive; (ii) direct van der Waals interactions between lipid chains $P_{\text{vdW}}$ are negligible; (iii) the direct and indirect electrostatic interactions ($P_{\rm el}$) between the charged lipid heads, mediated by counterions, are described to a first approximation by the interaction between two charge surfaces (charge density $\sigma_{\rm S}=-e/A_{\rm l}$), in the weak coupling regime (WC) or in the strong coupling regime (SC) depending on the experimental conditions (see below) \cite{Vishnyakov2017Nov}; (iv) indirect interactions mediated by water molecules are described by hydration repulsion $P_{\rm h}(z)=P_{\rm h0}e^{-z/z_{\rm h}}$ \cite{RAND1989351,Petrache(1998),Kowalik2017}; (v) finally, as confirmed by experiments, layer fluctuations are negligible (roughness $< 0.5$ nm) and surfaces can be considered as planes. In this context, the equilibrium position between the two lipid layers is given by solving the equation
\begin{equation}
P_{\rm{O}}+P_{\text{h}}+P_{\rm{el}}=0.
\label{eq_pressure_total}
\end{equation}
More details on the model are given in Supplementary Information (Section 6).

Finally, we must also take into account the fact that PC and PS heads have different hydration interaction parameters. Motivated by the work of Pincet et al.\cite{Xu2010Mar}, we attempt to describe the behavior of our system using a model that integrates the hydration repulsion of lipid mixtures. Considering that each phospholipid head interacts specifically with water and contributes independently and additively to the total hydration, we can express the hydration repulsion as a function of \(j\)

\begin{equation}
    P_{\text{h,tot}} = P_{\text{h0},\text{tot}} \exp \left( -\frac{z_{\rm cc}}{z_{\text{h},\text{tot}}} \right),
    \label{P_htot}
\end{equation}
where $P_{\text{h0,tot}} = P_{\text{h0,PC}}^{\left( (1 - j) z_{\text{h,PC}}/{z_{\text{h,tot}}} \right)} P_{\text{h0,PS}}^{\left( j  z_{\text{h,PS}}/{z_{\text{h,tot}}} \right)}$ and $z_{\text{h,tot}} = (1 - j)  z_{\text{h,PC}} + j  z_{\text{h,PS}}$. \(P_{\text{h0,PS}}\), \(z_{\text{h,PS}}\), \(P_{\text{h0,PC}}\), and \(z_{\text{h,PC}}\) are hydration constants related to DPPS and DPPC, respectively.

\begin{figure}[h!]
    \centering
    \includegraphics[width=\textwidth]{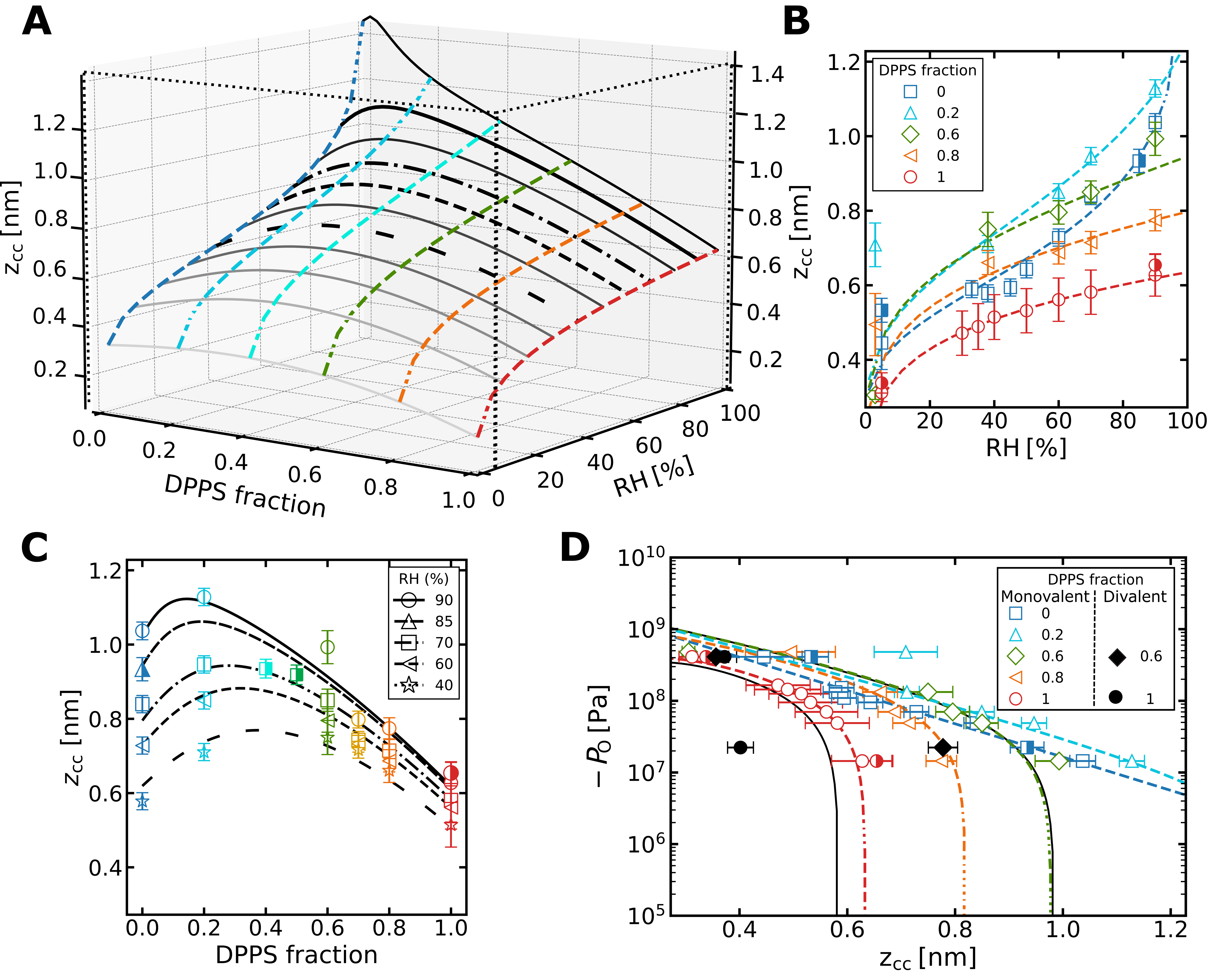}  
    \caption{(A) 3D plot showing model predictions, for monovalent counterions, of the distance between the effective charged surfaces of the lipid heads $z_{\rm cc}$ (solving equation \ref{eq_pressure_total}) as a function of DPPS fraction $j$ (colour code) and relative humidity RH. (B) and (C) Projections of the 3D plot (A) showing comparison between experimental data and best fits as a function of RH for 5 DPPS fractions (B) and as function of DPPS fraction at constant RH (C). Open symbols correspond to NR and semi-closed symbols to XRR. (D) Absolute value of osmotic pressure versus $z_{\rm cc}$, with the same color code as (B). Black filled symbols corresponds to experiments with divalent counterions (Ca$^{2+}$) and solid black lines to the model for a valence $q=2$. The best fits obtained by taking a confinement-independent permittivity $\epsilon_{\rm r}$ (dashed lines) or by calculating $\epsilon_{\rm r}$ from equation 9 (solid lines) are shown in (B) and (D).}
    \label{fig:3}
\end{figure}

\subsection{Pure DPPC layers ($j=0$)}

For pure DPPC (\(j=0\)), electrostatic interactions are negligible and \(z_{\rm cc}\) can be modeled using only osmotic and hydration pressures. Writing
$-P_{\rm{O}}=P_{\text{h}}$, we obtain a direct relationship between RH and \(z_{\rm cc}\). From the fit in Figure \ref{fig:3}, we extracted hydration pressure constants \(P_{\rm{h}0,\rm{PC}} = \left(3.6 \pm 1.0\right)\times 10^{9} \, \rm{Pa}\) and \(z_{\rm{h,\rm{PC}}} = 0.21 \pm 0.02\, \rm{nm}\), which are consistent with literature values for DPPC in the gel phase \cite{Kowalik2017}.

\subsection{Charged layers ($j\neq 0$)}

For pure DPPS \(j=1\), the significant reduction in $z_{\rm cc}$ for DPPS compared to DPPC (see Figures \ref{fig:2} and \ref{fig:3}) suggests the presence of SC interactions, as previously observed in double DPPS bilayers \cite{Mukhina2019Nov}. Interestingly, when examining $z_{\rm cc}$ variations between $j=0$ and $j=1$, an initial increase occurs at $j=0.2$, followed by a decrease up to $j=1$. This behavior is more clearly depicted in Figure \ref{fig:3}C, which shows $z_{\rm cc}$ variations with changing DPPS fraction at fixed RH. It is tempting to associate this bell curve shape with the signature of a crossover between the WC and SC regimes for electrostatic interactions ($P_{\rm el}$). To determine whether a system falls in the WC or SC regime, the coupling parameter 

\begin{equation}
\Xi = \frac{q^2 \, l_{\text{B}}}{\mu} \propto \frac{q^3 \sigma_{\rm{S}}}{\epsilon_{\rm r}^2},
\label{coupling-parameter}
\end{equation}
is calculated, where $q$ is the valence of the ions,  $l_{\text{B}} = e^2/(4 \pi \epsilon_0 \epsilon_{\rm r} k_{\text{B}} T)$ is the Bjerrum length\cite{bjerrum1926kgl}, and $\mu = 2\epsilon_0 \epsilon_{\rm r} k_{\text{B}} T/(q e \sigma_{\text{S}})$ is the Gouy-Chapman length\cite{gouy1910,Chapman1913}. This parameter quantifies the competition between counterion-counterion correlations and thermal agitation. For $\Xi < 12$, the coupling between ions is low, leading to the WC regime characterized by a repulsion, while for $\Xi \gg 12$, correlations cannot be neglected, setting the SC attraction \cite{Netz2001Aug}. The key to understand monovalent counterion electrostatic attraction lies in the fact that $\Xi$ scales inversely with the square of the permittivity. 

\begin{figure}[h]
\centering
\includegraphics[width=0.45\textwidth]{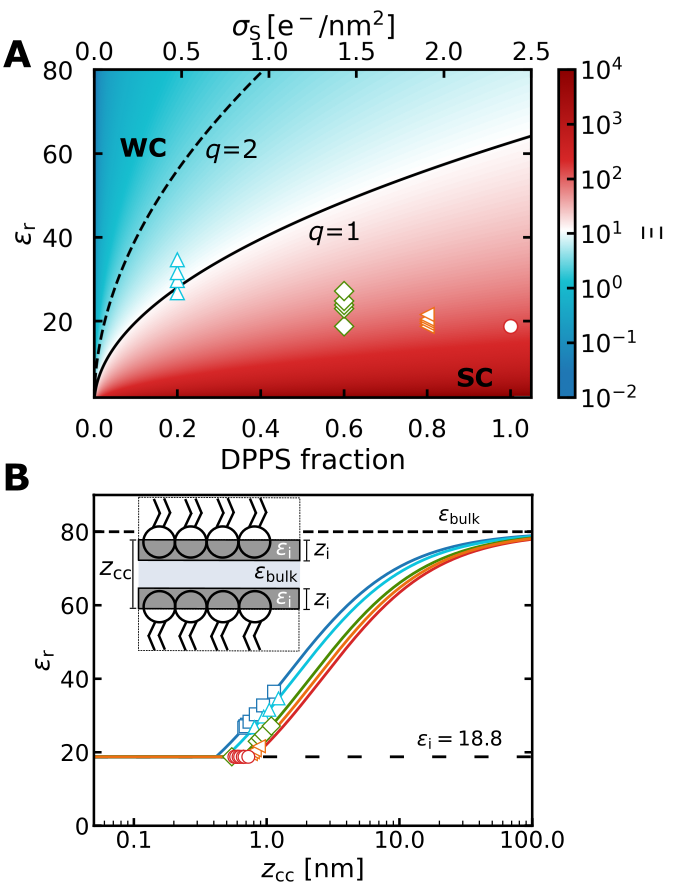}
\caption{(A) Color-coded projection in 2D of the  variation of the coupling parameter $\Xi$ as a function of the dielectric constant $\epsilon_{\rm r}$ and the DPPS fraction $j$. The corresponding surface charge density $\sigma_{\text{S}}$ is shown on the top x-axis. The solid line represents $\Xi=12$, marking the transition between the WC regime and the SC regime \cite{Moreira2002May} for monovalent counterions ($q=1$). The dashed line indicates the boundary for divalent counterions ($q=2$). (B) Dielectric constant $\epsilon_{\rm r}$ as a function of $z_{\rm cc}$ calculated with the equation \ref{eq_epsilon_mod} using the parameters deduced from the fits of the experimental results. For both (A) and (B) the color-coded symbols represent the experimental data for different surface charge densities $\sigma_{\rm S}$ (see Figure \ref{fig:3}), on which $\epsilon_{\rm r}$ depends through the constraint $z_{\rm i}=\lambda_{\rm h,tot}$.}
\label{fig:4}
\end{figure}

In Figure \ref{fig:4}, we illustrate how $\Xi$ varies with the surface charge and $\epsilon_{\rm r}$ ranging from bulk to optical values. If $\epsilon_{\rm r}$ indeed decreases toward the optical limit ($\epsilon_{\rm r}\sim 2.1$), as measured experimentally and numerically in recent papers \cite{Fumagalli2018Jun,Palaia2022Apr,borgis2023}, it is clear from Figure \ref{fig:4} that all charged mixtures discussed in this paper would already fall in the SC regime. We also compared the value of WC pressure $P_{\rm{WC}} = (k_{\rm{B}}T)^2 \pi \epsilon_0 \epsilon_{\rm{r}}/ (2 e^2 z_{\rm cc}^2)$ to the osmotic pressure\cite{ANDELMAN1995603,NAJI2005131}. Even at the highest humidity studied in this paper, WC pressure is at least 2 orders of magnitudes lower, discarding the possibility of WC pressure contributing to the disjoining pressure balance. The electrostatic contribution to total pressure is thus dominated by the SC pressure when \(j>0\) (see also Supporting Information, Sections 6)
\begin{equation}
P_{\rm{el}}=P_{\rm{SC}}= 2 \pi l_{\text{B}}\sigma_{\rm{S}}^2 k_{\rm B} T \left( \frac{2 \mu}{z_{\rm cc}} - 1 + \frac{1}{3\Xi} \frac{z_{\rm cc}}{\mu}\right),
\label{P_SC}
\end{equation}
where $l_{\text B}$ and $\mu$ depend on $\epsilon_{\text r}$ and $\mu$ on the counter-ion valency $q$\cite{Naji2013Oct,NAJI2005131}.

The equilibrium distance is thus given by solving:
\begin{equation}
P_{\rm{O}}+P_{\rm{SC}}+P_{\text{h}}=0,
\label{eq_pressure_total_SC}
\end{equation}
where $P_{\text{h,tot}}$ and $P_{\rm{SC}}$ are given by equations \ref{P_SC} and \ref{P_htot}. 
$z_{\text{h,PC}}$, $P_{\text{h0,PC}}$ being given by fitting the case $j=0$ as described above, this set of equations depends on the 3 unknown parameters \(z_{\text{h,PS}}\), \(P_{\text{h0,PS}}\) and $\epsilon_{\rm r}$. 

We first try to analyze our data by assuming that the relative permittivity between the charged planes (hydrated lipid heads and second water layer) is independent of $z_{\rm cc}$. Using the pressure-distance representation, we performed a global fit on all our DPPS fractions with the remaining 3 fitting parameters: \(P_{\text{h0,PS}}\), \(z_{\text{h,PS}}\) and $\epsilon_{\rm r}$. Best fits are shown in Figures \ref{fig:3} as dashed lines and exhibit a good agreement with the data across all RH and DPPS fractions with $P_{\text{h0,PS}}=1.75 \pm 0.25$ MPa, $z_{\text{h,PS}}=0.58 \pm 0.05$ nm and $\epsilon_{\rm r}= 15 \pm 2$. Our fitted parameters for DPPS hydration constants indicate a longer decay length for DPPS compared to neutral DPPC, which is consistent with experimental and numerical observations \cite{Chen2010Aug,Schlaich2019Jan}. The indirect hydration repulsion decay length can be compared to an independent estimate obtained from our MD simulations. Indeed, Kanduč, Schlaich  and co-workers \cite{Kanduc2014,Schlaich2024Apr} have shown that the water polarization of MD simulations can be interpreted as an order-parameter in a mean-field Landau-Ginzburg model. This model describes hydration repulsion between surfaces based on the perturbation of solvent structure by lipid headgroups. We compared the water polarization profiles from our simulations with model predictions to extract a decay length for the indirect hydration repulsion $\lambda_{\rm H}$ (see Supporting Information, Section 3), and obtained a value of $\lambda_{\rm H}=0.425\pm 0.025$ nm, significantly lower than the value of $z_{\rm h}$ derived from experiments. It is important to note that $z_{\rm h}$ and $\lambda_{\rm H}$ do not refer to the same physical mechanisms, since $z_{\rm H}$, extracted from polarization profiles, necessarily includes a contribution from the charged surface\footnote{We thank the anonymous reviewer for pointing this out.}.

The assumption that permittivity does not depend on the confinement $z_{\rm cc}$ is strong, and we have tried to go one step further by adapting the approach from Fumagalli et al. \cite{Fumagalli2018Jun}, where only a few water molecules near the surface are considered "frozen" with a dielectric constant of $\epsilon_{\rm i}$, while the water farther from the surface retains its bulk value. Treating the system as three capacitors in series, we obtain an effective permittivity dependent on \(z_{\rm cc}\)
\begin{equation}
    \frac{1}{\epsilon_{\rm r}} = \frac{1}{\epsilon_{\rm i}}\frac{2z_{\text i}}{z_{\rm cc}} + \frac{1}{\epsilon_{\rm bulk}} \left(1 - \frac{2z_{\rm i}}{z_{\rm cc}}\right).
\label{eq_epsilon_mod}
\end{equation}

This set of equations depends on 4 fitting parameters: \(z_{\text{h,PS}}\), \(P_{\text{h0,PS}}\), $z_{\rm i}$ and $\epsilon_{\rm i}$. To further constrain the analysis of the experiments, we chose to set $z_{\rm h, PS}$ to the characteristic length of dipole orientation decay obtained from numerical simulations ($z_{\rm h, PS}=0.4$ nm, see Supplementary Informations, Section 3). We also made the strong assumption that the characteristic length of the indirect hydration repulsion $z_{\rm h, tot}$ and the thickness of the frozen water layer $z_{\rm i}$ are equal. In the end, we are left with just 2 fitting parameters $P_{\text{h0,PS}}$ and $\epsilon_{\rm i}$. Best fits are shown in Figures C and D as solid lines and exhibit an excellent agreement with the data across all RH and DPPS fractions with $P_{\rm h0,PS} = 2.3 \pm 0.5$ MPa and $\epsilon_{\rm i} = 19 \pm 2$, which is significantly lower than the bulk value, but consistent with previous reports for nanoconfined water systems\cite{Fumagalli2018Jun, borgis2023, gardre2025}. As expected, this model seems to better describe the experimental data at low confinement ($z_{\rm cc}\sim 1$ nm), even if the differences are very close to the experimental uncertainties. We also show in Supplementary information (section 7) that the results obtained are not very sensitive to the choice of the position of the charged planes ($z_{\sigma_{\rm S}}$).

Using equation \ref{eq_epsilon_mod}, we calculated the dielectric permittivity values corresponding to the best-fit parameters. These are shown in Figure 4B as a function of the distance $z_{\rm cc}$. Some experimental data points fall in a range where the dielectric permittivity model is not strictly valid, specifically when $z_{\rm{cc}} < 2z_{\rm{i}}$. We still applied the model for continuity, a choice supported by Borgis et al. \cite{borgis2023}, who demonstrated its reliability even in this regime.

Finally, we also present in Figure 3D the results obtained with divalent counterions (Ca$^{2+}$) (black solid symbols). As expected, we observe a stronger attraction in the presence of divalent counterions. To assess this more quantitatively, we compared the data with the theoretical predictions obtained with the same fitting parameters, modifying only the counterion valency in the SC pressure (equation \ref{P_SC}, black curves). At high humidity, the agreement is not perfect, but the available data are insufficient to determine whether the SC model underestimates the electrostatic attraction, or whether other parameters, as for example ($z_{\rm{h0, PS}}$ and $P_{\rm h0, PS}$) are modified in the presence of calcium.


In conclusion, we measured experimentally the distance between lipid layers for different charge fractions, osmotic pressures and in the presence of monovalent or divalent counterions. Our experimental results unambiguously demonstrate the existence of electrostatic attraction between similarly charged layers, even in the presence of monovalent counterions. A continuous theoretical model, taking into account the hydration repulsion of lipid head mixtures and a strong electrostatic coupling between charged surfaces, enhanced by the reduced electrical permeability of confined water, allows us to describe all our experimental results in a unified way for different charge and relative humidity ratios using only three fitting parameters: the hydration pressure $P_{\rm h0,PS}$ and hydration length $z_{\rm h,PS}$ of DPPS molecules, and the dielectric constant of confined water. The results obtained for these three parameters are strongly supported by our numerical simulations.

This study provides a comprehensive analysis of the interplay between SC attraction and hydration repulsion in nanoconfined water between charged phospholipid layers. Our experimental results reveal that hydration water significantly reduces dielectric screening, leading to conditions where SC attraction becomes significant even in the absence of multivalent counterions. The balance between SC attraction and hydration repulsion governs the structural behavior of charged phospholipid layers. These findings enhance our understanding of the fundamental forces at play in biological and colloidal systems, particularly under conditions of nanoconfinement. 

\begin{acknowledgement}

The authors acknowledge the ANR – FRANCE (French National Research Agency) for its financial support of the AAPG2021 project n°ANR-21-CE30-0026. The authors thank Camille Robert, Lucas Saccucci for their help during the NR experiments, and Nicolas Aubert for technical assistance at SOLEIL. The authors thank Wiebke Drenckhan, Li Fu and Damien Favier for fruitful discussions. Awarded beamtimes at the ILL (10.5291/ILL-DATA.EASY-341) and at the SOLEIL (experiment 20250545) synchrotron are gratefully acknowledged. Support from PSCM facility at the ILL for sample preparation is gratefully acknowledged. This project was provided with computing HPC and storage resources by GENCI at IDRIS thanks to the grant 2023-A0130807662 on the supercomputer Jean Zay's V100 partition.

\end{acknowledgement}

\newpage
%
%

\begin{suppinfo}

Supporting Information Available: Detailed description of: (1) sample preparations; (2) X-ray and neutron reflectivities experiments and data analysis; (3) numerical simulations; (4) X-ray fluorescence experiments; (5) charge location of charged planes;  (6) theoretical model and (7) influence of $z_{\sigma_{\rm S}}$ on fitted parameters.

\end{suppinfo}





\providecommand{\latin}[1]{#1}
\makeatletter
\providecommand{\doi}
  {\begingroup\let\do\@makeother\dospecials
  \catcode`\{=1 \catcode`\}=2 \doi@aux}
\providecommand{\doi@aux}[1]{\endgroup\texttt{#1}}
\makeatother
\providecommand*\mcitethebibliography{\thebibliography}
\csname @ifundefined\endcsname{endmcitethebibliography}
  {\let\endmcitethebibliography\endthebibliography}{}

  
\appendix

\newpage

\pagestyle{empty}



\section*{Supplementary material}

# Supplementary Information for
# The role of confined water in the emergence of electrostatic strong coupling as revealed by nanoseparated charged lipid layers

Swen Helstroffer[1], Ludovic Gardré[2], Giovanna Fragneto[3], Arnaud Hemmerle[4],
Léo Henry[1], Laurent Joly[2], Fabrice Thalmann[1], Claire Loison[2], Pierre Muller[1], Thierry Charitat[1]

May 16, 2025

[1]Université de Strasbourg, CNRS, Institut Charles Sadron UPR22, F-67000 Strasbourg, France
[2]Université Claude Bernard Lyon 1, CNRS, Institut Lumière Matière, UMR5306, F69100 Villeurbanne, France
[3]Institut Laue-Langevin, 71 av. des Martyrs, BP 156, 38042 Grenoble Cedex, France (current address: European Spallation source ERIC, Partikelgatan 2, P.O. Box 176, SE-221 00 Lund, Sweden)
[4]Synchrotron SOLEIL, L'Orme des Merisiers, Départementale 128, 91190 Saint-Aubin, France

## 1 Sample preparations

### 1.1 Langmuir-Blodgett deposition

Samples were prepared using the Langmuir-Blodgett (LB) deposition technique **??**. Neutral phospholipids are generally limited to three layers **?**. For anionic DPPS phospholipids and certain DPPS/DPPC mixtures, it was possible to deposit more layers.

### 1.2 Substrate cleaning

The silicon substrates were atomically smooth polished **?** with varying sizes depending on the experiment. To ensure a highly hydrophilic substrate free from organic residues, the standard cleaning method involves ultrasonic baths of solvents from less polar to more polar: chloroform, acetone, ethanol, and Milli-Q water. The substrate is then dried with nitrogen gas and exposed to UV/ozone or plasma treatment before a final rinse with Milli-Q water. Immediately afterward, the silicon substrate is immersed in the Langmuir-Blodgett trough.

### 1.3 Lipid solution and counterions

Phospholipid solutions were prepared at 1 mg/ml in chloroform for DSPC and DPPC. DPPS, which is less soluble in chloroform, required a solvent mixture of chloroform:methanol:water at a ratio of 70:29:1 to fully solubilize it at 1 mg/ml. This solvent ratio was used for pure DPPS and all DPPS/DPPC mixtures. Proper solubilization is essential for accurate phospholipid transfer to the air/water interface. Sometimes, mild heating and sonication of the DPPS solution were necessary for homogenization.

## 1.4 Langmuir isotherm

We used a KSV NIMA Langmuir-Blodgett Deposition trough **?**. The substrate was first dipped into the subphase (Milli-Q water with or without added salt, see below). The lipid solution was added drop-by-drop to the water surface using a Hamilton syringe. Slow, controlled phospholipid solution deposition ensures a reproducible monolayer formation, as the solvent evaporates in a few tens of minutes, leaving the phospholipids at the water surface.

Isotherms were performed at a constant temperature of 25°C, maintained by a thermostated bath. By gradually compressing the layers (compression speed of 10 mm/min), we recorded isotherms for pure DPPC, DPPS, DSPC and for DPPC/DPPS mixtures.

## 1.5 Langmuir-Blodgett deposition

For the transfer of the lipid layers onto the substrate, the surface pressure was maintained at constant value by the KSV system. The substrate is pulled out of (or pushed into) the subphase (at a constant speed of 4 mm/min), the monolayer deposits onto the substrate. As the monolayer transfers, the barriers are moved to maintain the same density at the air-water interface and compensate for the molecules transferred onto the solid surface. The change in the trough area while moving the substrate allows to evaluate the instantaneous transfer rate (see Figure S1).

### 1.5.1 First DSPC monolayer

A first monolayer of DSPC is deposited on the substrate at a constant pressure of $\Pi = 40$ mN/m.

### 1.5.2 Second and third monolayers-Counterions exchange

This work focuses on the interactions between the second and third lipid layers. To study the effect of the surface charge of these two layers, the following deposits were made from monolayers consisting of a mixture of DPPCs and DPPSs of varying compositions. We also varied the nature of the counterions present between the lipid layers. To this end, counterions were exchanged using ionic solutions of Caesium hydroxide (CsOH) or Calcium hydroxide ((Ca(OH)$_2$) at $10^{-2}$ M as a subphase for the deposition of the second and third layer.

After creating a new isotherm using the DPPS/DPPC mixture of the desired composition and the appropriate subphase, the substrate is dipped again into the subphase, resulting in the formation of a bilayer at a constant surface pressure of 40 or 45 mN/m. After the deposition of the third layer, we obtain a trilayer exposed to air, as shown in Figure S1. For DSPC and DPPC, this represents the limit of the number of layers that can be stacked by classical LB deposition. If an attempt is made to deposit a fourth layer, the third layer is re-transferred onto the air-water interface (see Figure S1, blue dashed line).

### 1.5.3 More than 3 layers

In contrast, for DPPS and DPPS/DPPC mixtures with a mass fraction of DPPS greater than $j = 0.6$ (where $j = m_{\text{DPPS}}/(m_{\text{DPPS}} + m_{\text{DPPC}})$), it is possible to stack an indefinite number of layers **?**; in our experiments, we limit the stacking to five layers, referred to as pentalayers.

### 1.5.4 Transfer rates

To quantify the quality of the deposition, a good approach is to compute the average transfer ratio TR defined as:

$$\text{TR} = \frac{\Delta A_{\text{trough}}}{\Delta A_{\text{substrate}}}. \tag{S1}$$

If TR is close to one, it indicates that the molecular density on the solid substrate is similar to the one at the air-water interface. Typically, the transfer ratio is slightly less than one. Figure S1 compares the transfer rate curves for a DPPS pentalayer and a DSPC trilayer deposition, while table S1 compares the average TR values.

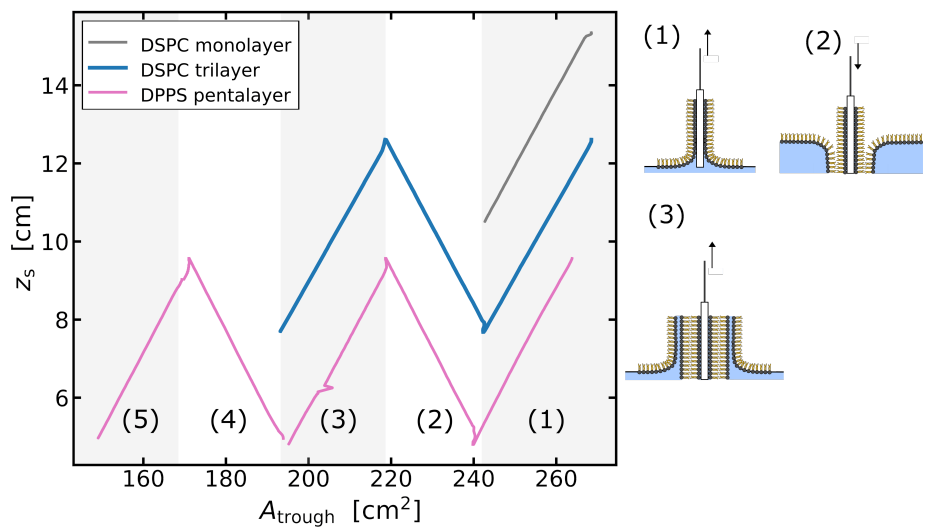

Figure S1: Transfer rate curves representing the position of the substrate $z_S$ as a function of the area of the barrier. Black (—, top) for a monolayer of DSPC. In blue (—, middle) for a trilayer of DSPC. In pink (—, bottom), deposition of a pentalayer of DPPS.

| Sample | Layer 1 | Layer 2 | Layer 3 | Layer 4 | Layer 5 | Subphase |
|---|---|---|---|---|---|---|
| 1 | $j = 0$ | $j = 0$ | $j = 0$ | – | – | Pure water |
|   | 0.97 | 0.97 | 0.96 | – | – | |
| 2 | $j = 0$ | $j = 1$ | $j = 1$ | – | – | Pure water |
|   | 1.16 | 0.98 | 0.99 | – | – | |
| 3 | $j = 0.2$ | $j = 0.2$ | $j = 0.2$ | – | – | Pure water |
|   | 0.96 | 0.97 | 0.99 | – | – | |
| 4 | $j = 0$ | $j = 0.6$ | $j = 0.6$ | – | – | Pure water |
|   | 0.99 | 0.85 | 1.00 | – | – | |
| 5 | $j = 0$ | $j = 0.7$ | $j = 0.7$ | $j = 0.7$ | $j = 0.7$ | Pure water |
|   | 1.01 | 0.60 | 0.97 | 0.82 | 0.96 | |
| 6 | $j = 0$ | $j = 0.8$ | $j = 0.8$ | $j = 0.8$ | $j = 0.8$ | Pure water |
|   | 1.00 | 0.61 | 0.96 | 1.02 | 1.04 | |

Table S1: Lipid composition (expressed as mass fraction $j$ of DPPS) and transfer rate for NR samples.

| Sample | Layer 1 | Layer 2 | Layer 3 | Subphase |
|---|---|---|---|---|
| 7 | DSPC 0.98 | $j = 0$ 0.96 | $j = 0$ 0.99 | Pure water |
| 8 | DSPC 1.01 | $j = 1$ 0.91 | $j = 1$ 0.98 | Pure water |
| 9 | DSPC 1.00 | $j = 0.4$ 0.99 | $j = 0.4$ 1.00 | Pure water |
| 10 | DSPC 0.66 | $j = 0.5$ 0.92 | $j = 0.5$ 1.03 | CsOH 0.1 mM |
| 11 | DSPC 0.95 | $j = 0.5$ 0.88 | $j = 0.5$ 1.01 | CaCl$_2$ 0.05 mM |
| 12 | DSPC 1.01 | $j = 1$ 0.92 | $j = 1$ 0.95 | CaCl$_2$ 0.1 mM |

Table S2: Lipid composition (expressed as mass fraction $j$ of DPPS) and transfer rate for XRR samples. The subphase used during deposition is specified for each sample; note that the first layer (DSPC) was always deposited from pure water.

### 1.5.5 Influence of ambient humidity

We wanted to emphasize the influence of RH on monolayer deposition. As far as we know, it has not been previously mentioned in the literature.

We have strong evidence that maintaining a low RH during Langmuir-Blodgett deposition strongly improves deposition quality. Although we do not have a quantitative analysis of this effect, we can compare the deposition of DSPC trilayers without RH control under poor conditions to those where we tuned the RH. In the latter case, we injected dry air (RH=3%) into the laminar flow hood containing the Langmuir trough, allowing to achieve a low RH value of about 30%.

Since we began controlling the RH, we did not experience any failed depositions, which were fairly frequent before. A plausible explanation is that low RH corresponds to strong osmotic pressure. This osmotic pressure stabilizes the layer and prevents re-transfer to the water interface. Dipping the substrate vertically into the water may induce shear on the already deposited layer. The osmotic pressure would help to "glue" the confined layers together. This concept of binding layers also applies to anionic DPPS phospholipid layers, where we can stack more layers.

## 2 Reflectivity Analysis

Reflectivity analysis is inherently model-dependent. In our case, experimental data are interpreted using a classical slab model. Our system, composed of stacked phospholipid layers, is particularly well-suited for this approach. Each part of the lamellar sample is characterized by thickness, scattering length density (SLD), and an inter-slab roughness. These parameters define the SLD profile, which is then related to the reflectivity, $R(q_z)$, through the Abeles method **?**.

The NR and XRR data were analyzed using the refnx Python package **?**. A 10-box model was used for NR trilayers, and a 15-box model for pentalayers. For high-resolution XRR data, an additional layer representing the central methyl group between the two leaflets was included, resulting in an 11-box model. The analysis was performed with the following constraints:

1. SLDs are fixed and calculated from theoretical volumes (see tables S3 and S4);

2. Thickness, roughness and solvent fraction of the silica layer ($SiO_2$) are previously adjusted by analysis of a bare substrate;

3. Chain and head parameters (thicknesses, solvent fractions and roughnesses) of the 2nd and 3rd lipid layers (trilayer) or the 2nd, 3rd, 4th and 5th lipid layers (penta-layer) are considered identical in the following fits;

4. The thickness of the phospholipid heads, denoted as $d_{\text{head}}$, are calculated from the fitted length of the chain $d_{\text{tail}}$ and using the theoretical volumes of the tail $V_{\text{m, tail}}$ and of the head $V_{\text{m, head}}$ as:

$$d_{\text{head}} = \frac{V_{\text{m, head}} \cdot d_{\text{tail}}}{V_{\text{m,tail}}}.$$

NR and XRR experiments were performed in humidity-controlled chambers, allowing for precise adjustment of relative humidity (RH). The goal of these experiments was to investigate the evolution of phospholipid headgroup hydration at different RH values (or the corresponding osmotic pressures). NR measurements were conducted in the presence of both $H_2O$ and $D_2O$ vapor, enabling co-refinement of the reflectivity profiles.

Table S3: Neutron scattering length density of used phospholipids

| Name | Structure | $b_n$ | $V_m$ [nm$^3$] | SLDn [$10^{-4}$ nm$^{-2}$] |
|---|---|---|---|---|
| PC head | $C_{10}H_{18}O_8PN$ | 60.054 | 0.319 | 1.88 |
| PS head | $C_8H_{11}O_{10}PNa$ | 81.55 | 0.31 | 2.63 |
| DS tail | $C_{34}H_{70}$ | -35.836 | 0.917 | -0.39 |
| DP tail | $C_{30}H_{62}$ | -32.5 | 0.825 | -0.39 |

Table S4: X-ray scattering length density of used phospholipids

| Name | Structure | $Z_{el}$ | $V_m$ [nm$^3$] | SLDx [10$^{-4}$ nm$^{-2}$] |
|---|---|---|---|---|
| PC head | $C_{10}H_{18}O_8PN$ | 164 | 0.319 | 14.45 |
| PS head | $C_8H_{11}O_{10}PNa$ | 172 | 0.31 | 15.5 |
| DS tail | $C_{34}H_{68}$ | 272 | 0.917 | 8.2 |
| DP tail | $C_{30}H_{60}$ | 240 | 0.825 | |
| CH$_3$ | CH$_3$ | 9 | 0.038 | 6.65 |

Table S5: Scattering length density of solvents and substrate. The SLD$_x$ values for Si and SiO$_2$ are not fixed; the values shown here correspond to their initial parameters before minimization.

| Material | SLDn [10$^{-4}$ nm$^{-2}$] | SLDx [10$^{-4}$ nm$^{-2}$] |
|---|---|---|
| H$_2$O | -0.56 | 9.41 |
| D$_2$O | 6.36 | - |
| Si | 2.07 | 18 |
| SiO$_2$ | 3.41 | 19 |

## 2.1 Neutron Reflectivity

### 2.1.1 Sample 1: DPPC-DPPC-DPPC ($j = 0$)

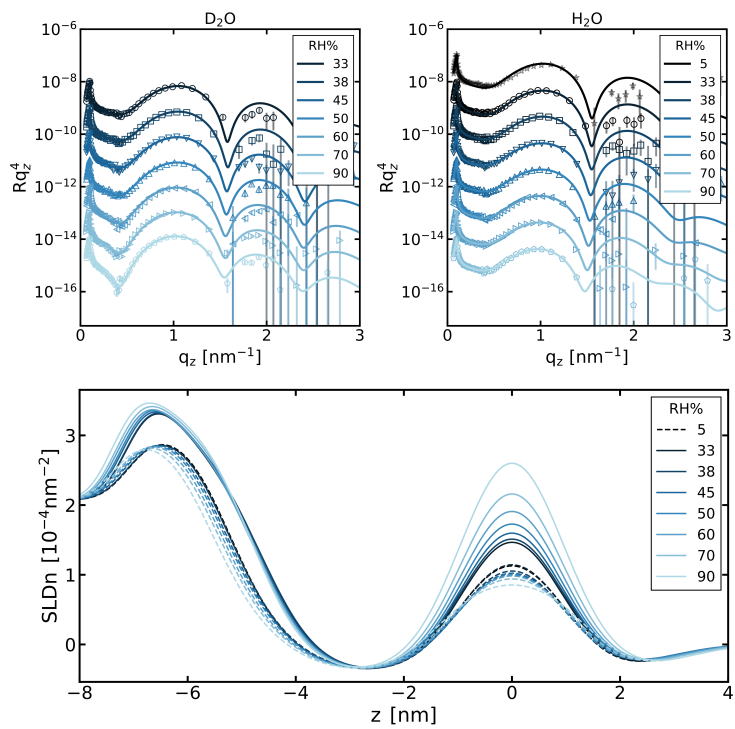

Figure S2: (Top) NR data of pure DPPC trilayer (Sample 1) for different relative humidity. Data are presented as $Rq_z^4$ on a semilog scale with two contrasts: (Left) in $D_2O$ and (Right) in $H_2O$. Solid lines represent the best joined fits. (Bottom) $SLD_n$ profiles corresponding to the fits from above in $D_2O$ (Solid lines) and in $H_2O$ (Dashed line).

Table S6: Sample 1, parameters at RH = 5%

| # | Slab | Thickness (nm) | SLD ($\times 10^{-4}$ nm$^{-2}$) | Roughness (nm) | vfsolv (%) |
|---|---|---|---|---|---|
| 0 | air | 0.00 | 0.00 | 0.00 | 0.00 |
| 1 | tail | 2.06 | -0.39 | 0.77 | 0.00 |
| 2 | head $j = 0$ | 0.80 | 1.88 | 0.77 | 0.00 |
| 3 | solvent | 0.11 ± 0.03 | -0.56 | 0.77 | 0.00 |
| 4 | head $j = 0$ | 0.80 | 1.88 | 0.77 | 0.00 |
| 5 | tail | 2.06 | -0.39 | 0.77 | 0.00 |
| 6 | tail | 1.80 | -0.39 | 0.89 | 0.00 |
| 7 | head $j = 0$ | 0.70 | 1.88 | 0.89 | 0.00 |
| 8 | solvent | 0.00 ± 0.02 | -0.56 | 0.89 | 0.00 |
| 9 | SiO$_2$ | 1.69 | 3.41 | 0.69 | 0.13 |
| 10 | Si | 0.00 | 2.07 | 0.39 | 0.00 |

Table S7: Sample 1, parameters at RH = 33%

| # | Slab | Thickness (nm) | SLD ($\times 10^{-4}$ nm$^{-2}$) | Roughness (nm) | vfsolv (%) |
|---|---|---|---|---|---|
| 0 | air | 0.00 | 0 | 0.00 | 0.00 |
| 1 | tail | 2.06 ± 0.12 | -0.39 | 0.77 ± 0.11 | 0.00 |
| 2 | head $j=0$ | 0.80 | 1.88 | 0.77 ± 0.11 | 0.00 |
| 3 | solvent | 0.09 ± 0.03 | 6.36/-0.56 | 0.77 ± 0.11 | 0.00 |
| 4 | head $j=0$ | 0.80 | 1.88 | 0.77 ± 0.11 | 0.00 |
| 5 | tail | 2.06 ± 0.12 | -0.39 | 0.77 ± 0.11 | 0.00 |
| 6 | tail | 1.80 ± 0.16 | -0.39 | 0.89 ± 0.06 | 0.00 |
| 7 | head $j=0$ | 0.70 | 1.88 | 0.89 ± 0.06 | 0.00 |
| 8 | solvent | 0.02 ± 0.06 | 6.36/-0.56 | 0.89 ± 0.06 | 0.00 |
| 9 | SiO2 | 1.69 | 3.41 | 0.69 | 0.13 |
| 10 | Si | 0.00 | 2.07 | 0.39 | 0.00 |

Table S8: Sample 1, parameters at RH = 38%

| # | Slab | Thickness (nm) | SLD ($\times 10^{-4}$ nm$^{-2}$) | Roughness (nm) | vfsolv (%) |
|---|---|---|---|---|---|
| 0 | air | 0.0 | 0.00 | 0.0 | 0.0 |
| 1 | tail | 2.0 ± 0.01 | -0.39 | 0.80 ± 0.01 | 0.0 |
| 2 | head $j = 0$ | 0.8 | 1.88 | 0.80 ± 0.01 | 0.0 |
| 3 | solvent | 0.13 ± 0.00 | 6.36 / -0.56 | 0.80 ± 0.01 | 0.0 |
| 4 | head $j = 0$ | 0.8 | 1.88 | 0.80 ± 0.01 | 0.0 |
| 5 | tail | 2.0 ± 0.01 | -0.39 | 0.80 ± 0.01 | 0.0 |
| 6 | tail | 1.8 ± 0.02 | -0.39 | 0.90 ± 0.01 | 0.0 |
| 7 | head $j = 0$ | 0.7 | 1.88 | 0.90 ± 0.01 | 0.0 |
| 8 | solvent | 0.04 ± 0.01 | 6.36 / -0.56 | 0.90 ± 0.01 | 0.0 |
| 9 | SiO$_2$ | 1.7 | 3.41 | 0.7 | 0.13 |
| 10 | Si | 0.0 | 2.07 | 0.4 | 0.0 |

Table S9: Sample 1, parameters at RH = 45%

| # | Slab | Thickness (nm) | SLD ($\times 10^{-4}$ nm$^{-2}$) | Roughness (nm) | vfsolv (%) |
|---|------|----------------|------------------------------------|----------------|------------|
| 0 | air | 0.0 | 0.00 | 0.0 | 0.0 |
| 1 | tail | 2.0 ± 0.01 | -0.39 | 0.80 ± 0.01 | 0.0 |
| 2 | head $j = 0$ | 0.8 | 1.88 | 0.80 ± 0.01 | 0.0 |
| 3 | solvent | 0.17 ± 0.00 | 6.36 / -0.56 | 0.80 ± 0.01 | 0.0 |
| 4 | head $j = 0$ | 0.8 | 1.88 | 0.80 ± 0.01 | 0.0 |
| 5 | tail | 2.0 ± 0.01 | -0.39 | 0.80 ± 0.01 | 0.0 |
| 6 | tail | 1.9 ± 0.02 | -0.39 | 0.89 ± 0.01 | 0.0 |
| 7 | head $j = 0$ | 0.7 | 1.88 | 0.89 ± 0.01 | 0.0 |
| 8 | solvent | 0.05 ± 0.01 | 6.36 / -0.56 | 0.89 ± 0.01 | 0.0 |
| 9 | SiO$_2$ | 1.7 | 3.41 | 0.7 | 0.13 |
| 10 | Si | 0.0 | 2.07 | 0.4 | 0.0 |

Table S10: Sample 1, parameters at RH = 50%

| # | Slab | Thickness (nm) | SLD ($\times 10^{-4}$ nm$^{-2}$) | Roughness (nm) | vfsolv (%) |
|---|---|---|---|---|---|
| 0 | air | 0.0 | 0.00 | 0.0 | 0.0 |
| 1 | tail | 2.0 ± 0.01 | -0.39 | 0.80 ± 0.01 | 0.0 |
| 2 | head $j = 0$ | 0.8 | 1.88 | 0.80 ± 0.01 | 0.0 |
| 3 | solvent | 0.21 ± 0.00 | 6.36 / -0.56 | 0.80 ± 0.01 | 0.0 |
| 4 | head $j = 0$ | 0.8 | 1.88 | 0.80 ± 0.01 | 0.0 |
| 5 | tail | 2.0 ± 0.01 | -0.39 | 0.80 ± 0.01 | 0.0 |
| 6 | tail | 1.9 ± 0.02 | -0.39 | 0.89 ± 0.01 | 0.0 |
| 7 | head $j = 0$ | 0.7 | 1.88 | 0.89 ± 0.01 | 0.0 |
| 8 | solvent | 0.07 ± 0.01 | 6.36 / -0.56 | 0.89 ± 0.01 | 0.0 |
| 9 | SiO$_2$ | 1.7 | 3.41 | 0.7 | 0.13 |
| 10 | Si | 0.0 | 2.07 | 0.4 | 0.0 |

Table S11: Sample 1, parameters at RH = 60%

| # | Slab | Thickness (nm) | SLD ($\times 10^{-4}$ nm$^{-2}$) | Roughness (nm) | vfsolv (%) |
|---|---|---|---|---|---|
| 0 | air | 0.0 | 0.00 | 0.0 | 0.0 |
| 1 | tail | 2.01 ± 0.01 | -0.39 | 0.80 ± 0.01 | 0.0 |
| 2 | head $j = 0$ | 0.78 | 1.88 | 0.80 ± 0.01 | 0.0 |
| 3 | solvent | 0.27 ± 0.00 | 6.36 / -0.56 | 0.80 ± 0.01 | 0.0 |
| 4 | head $j = 0$ | 0.78 | 1.88 | 0.80 ± 0.01 | 0.0 |
| 5 | tail | 2.01 ± 0.01 | -0.39 | 0.80 ± 0.01 | 0.0 |
| 6 | tail | 1.84 ± 0.02 | -0.39 | 0.90 ± 0.01 | 0.0 |
| 7 | head $j = 0$ | 0.71 | 1.88 | 0.90 ± 0.01 | 0.0 |
| 8 | solvent | 0.09 ± 0.01 | 6.36 / -0.56 | 0.90 ± 0.01 | 0.0 |
| 9 | SiO$_2$ | 1.69 | 3.41 | 0.69 | 0.13 |
| 10 | Si | 0.0 | 2.07 | 0.39 | 0.0 |

Table S12: Sample 1, parameters at RH = 70%

| # | Slab | Thickness (nm) | SLD ($\times 10^{-4}$ nm$^{-2}$) | Roughness (nm) | vfsolv (%) |
|---|---|---|---|---|---|
| 0 | air | 0.00 | 0.00 | 0.00 | 0.0 |
| 1 | tail | 2.04 ± 0.01 | -0.39 | 0.80 ± 0.01 | 0.0 |
| 2 | head $j = 0$ | 0.79 | 1.88 | 0.80 ± 0.01 | 0.0 |
| 3 | solvent | 0.36 ± 0.00 | 6.36 / -0.56 | 0.80 ± 0.01 | 0.0 |
| 4 | head $j = 0$ | 0.79 | 1.88 | 0.80 ± 0.01 | 0.0 |
| 5 | tail | 2.04 ± 0.01 | -0.39 | 0.80 ± 0.01 | 0.0 |
| 6 | tail | 1.77 ± 0.02 | -0.39 | 0.89 ± 0.01 | 0.0 |
| 7 | head $j = 0$ | 0.69 | 1.88 | 0.89 ± 0.01 | 0.0 |
| 8 | solvent | 0.13 ± 0.01 | 6.36 / -0.56 | 0.89 ± 0.01 | 0.0 |
| 9 | SiO$_2$ | 1.69 | 3.41 | 0.69 | 0.13 |
| 10 | Si | 0.00 | 2.07 | 0.39 | 0.0 |

Table S13: Sample 1, parameters at RH = 90%

| # | Slab | Thickness (nm) | SLD ($\times 10^{-4}$ nm$^{-2}$) | Roughness (nm) | vfsolv (%) |
|---|---|---|---|---|---|
| 0 | air | 0.00 | 0.00 | 0.00 | 0.0 |
| 1 | tail | 2.09 ± 0.01 | -0.39 | 0.80 ± 0.01 | 0.0 |
| 2 | head $j = 0$ | 0.81 | 1.88 | 0.80 ± 0.01 | 0.0 |
| 3 | solvent | 0.51 ± 0.00 | 6.36 / -0.56 | 0.80 ± 0.01 | 0.0 |
| 4 | head $j = 0$ | 0.81 | 1.88 | 0.80 ± 0.01 | 0.0 |
| 5 | tail | 2.09 ± 0.01 | -0.39 | 0.80 ± 0.01 | 0.0 |
| 6 | tail | 1.67 ± 0.01 | -0.39 | 0.89 ± 0.01 | 0.0 |
| 7 | head $j = 0$ | 0.65 | 1.88 | 0.89 ± 0.01 | 0.0 |
| 8 | solvent | 0.19 ± 0.01 | 6.36 / -0.56 | 0.89 ± 0.01 | 0.0 |
| 9 | SiO$_2$ | 1.69 | 3.41 | 0.69 | 0.13 |
| 10 | Si | 0.00 | 2.07 | 0.39 | 0.0 |

### 2.1.2 Sample 2: DPPC-DPPS-DPPS ($j = 1$)

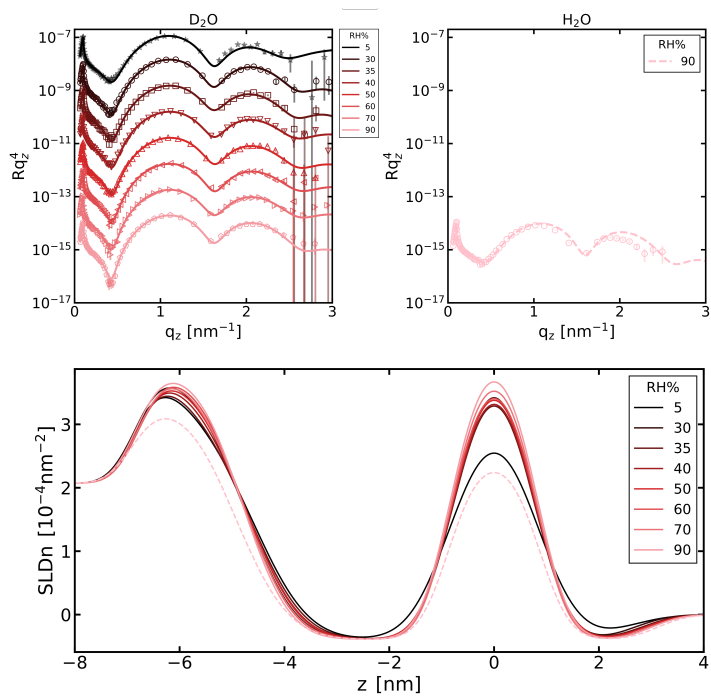

Figure S3: (Top) NR data of a DPPC-DPPS-DPPS trilayer (Sample 2) for different relative humidity. Data are presented as $Rq_z^4$ on a semilog scale in $D_2O$. (Bottom) $SLD_n$ profiles corresponding to the fits from above in $D_2O$.

16

Table S14: Sample 2, parameters at RH = 5%

| # | Slab | Thickness (nm) | SLD ($\times 10^{-4}$ nm$^{-2}$) | Roughness (nm) | vfsolv (%) |
|---|---|---|---|---|---|
| 0 | air | 0.00 | 0.00 | 0.00 | 0.0 |
| 1 | tail | 1.72 ± 0.02 | -0.39 | 0.65 ± 0.01 | 0.0 |
| 2 | head $j = 1$ | 0.80 ± 0.01 | 2.63 | 0.65 ± 0.01 | 0.19 ± 0.01 |
| 3 | solvent | 0.00 | 6.36 | 0.65 ± 0.01 | 0.0 |
| 4 | head $j = 1$ | 0.80 ± 0.01 | 2.63 | 0.65 ± 0.01 | 0.19 ± 0.01 |
| 5 | tail | 1.72 ± 0.02 | -0.39 | 0.65 ± 0.01 | 0.0 |
| 6 | tail | 1.92 ± 0.02 | -0.39 | 0.80 ± 0.00 | 0.0 |
| 7 | head $j = 0$ | 0.74 ± 0.01 | 1.88 | 0.80 ± 0.00 | 0.0 ± 0.0 |
| 8 | solvent | 0.00 ± 0.00 | 6.36 | 0.80 ± 0.00 | 0.0 |
| 9 | SiO$_2$ | 1.69 | 3.41 | 0.69 | 0.13 |
| 10 | Si | 0.00 | 2.07 | 0.39 | 0.0 |

Table S15: Sample 2, parameters at RH = 30%

| # | Slab | Thickness (nm) | SLD ($\times 10^{-4}$ nm$^{-2}$) | Roughness (nm) | vfsolv (%) |
|---|---|---|---|---|---|
| 0 | air | 0.00 | 0.00 | 0.00 | 0.0 |
| 1 | tail | 1.91 ± 0.08 | -0.39 | 0.52 ± 0.01 | 0.0 |
| 2 | head $j = 1$ | 0.72 | 2.63 | 0.52 ± 0.01 | 0.28 |
| 3 | solvent | 0.13 ± 0.04 | 6.36 | 0.52 ± 0.01 | 0.0 |
| 4 | head $j = 1$ | 0.72 | 2.63 | 0.52 ± 0.01 | 0.28 |
| 5 | tail | 1.91 ± 0.08 | -0.39 | 0.52 ± 0.01 | 0.0 |
| 6 | tail | 1.77 ± 0.07 | -0.39 | 0.69 ± 0.01 | 0.0 |
| 7 | head $j = 0$ | 0.69 | 1.88 | 0.69 ± 0.01 | 0.0 |
| 8 | solvent | 0.00 ± 0.00 | 6.36 | 0.69 ± 0.01 | 0.0 |
| 9 | SiO$_2$ | 1.69 | 3.41 | 0.69 | 0.13 |
| 10 | Si | 0.00 | 2.07 | 0.39 | 0.0 |

Table S16: Sample 2, parameters at RH = 35%

| # | Slab | Thickness (nm) | SLD ($\times 10^{-4}$ nm$^{-2}$) | Roughness (nm) | vfsolv (%) |
|---|---|---|---|---|---|
| 0 | air | 0.00 | 0.00 | 0.00 | 0.0 |
| 1 | tail | 2.03 ± 0.08 | -0.39 | 0.52 ± 0.01 | 0.0 |
| 2 | head $j = 1$ | 0.76 | 2.63 | 0.52 ± 0.01 | 0.28 |
| 3 | solvent | 0.06 ± 0.04 | 6.36 | 0.52 ± 0.01 | 0.0 |
| 4 | head $j = 1$ | 0.76 | 2.63 | 0.52 ± 0.01 | 0.28 |
| 5 | tail | 2.03 ± 0.08 | -0.39 | 0.52 ± 0.01 | 0.0 |
| 6 | tail | 1.67 ± 0.07 | -0.39 | 0.78 ± 0.01 | 0.0 |
| 7 | head PC $j = 0$ | 0.65 | 1.88 | 0.78 ± 0.01 | 0.0 |
| 8 | solvent | 0.00 ± 0.01 | 6.36 | 0.78 ± 0.01 | 0.0 |
| 9 | SiO$_2$ | 1.69 | 3.41 | 0.69 | 0.13 |
| 10 | Si | 0.00 | 2.07 | 0.39 | 0.0 |

Table S17: Sample 2, parameters at RH = 40%

| # | Slab | Thickness (nm) | SLD ($\times 10^{-4}$ nm$^{-2}$) | Roughness (nm) | vfsolv (%) |
|---|---|---|---|---|---|
| 0 | air | 0.00 | 0.00 | 0.00 | 0.0 |
| 1 | tail | 2.09 ± 0.08 | -0.39 | 0.51 ± 0.02 | 0.0 |
| 2 | head $j = 1$ | 0.79 | 2.63 | 0.51 ± 0.02 | 0.28 |
| 3 | solvent | 0.04 ± 0.04 | 6.36 | 0.51 ± 0.02 | 0.0 |
| 4 | head $j = 1$ | 0.79 | 2.63 | 0.51 ± 0.02 | 0.28 |
| 5 | tail | 2.09 ± 0.08 | -0.39 | 0.51 ± 0.02 | 0.0 |
| 6 | tail | 1.60 ± 0.07 | -0.39 | 0.74 ± 0.01 | 0.0 |
| 7 | head $j = 0$ | 0.62 | 1.88 | 0.74 ± 0.01 | 0.0 |
| 8 | solvent | 0.00 ± 0.01 | 6.36 | 0.74 ± 0.01 | 0.0 |
| 9 | SiO$_2$ | 1.69 | 3.41 | 0.69 | 0.13 |
| 10 | Si | 0.00 | 2.07 | 0.39 | 0.0 |

Table S18: Sample 2, parameters at RH = 50%

| # | Slab | Thickness (nm) | SLD ($\times 10^{-4}$ nm$^{-2}$) | Roughness (nm) | vfsolv (%) |
|---|------|----------------|----------------------------------|----------------|------------|
| 0 | air | 0.00 | 0.00 | 0.00 | 0.0 |
| 1 | tail | 2.08 ± 0.08 | -0.39 | 0.51 ± 0.01 | 0.0 |
| 2 | head $j = 1$ | 0.78 | 2.63 | 0.51 ± 0.01 | 0.28 |
| 3 | solvent | 0.07 ± 0.04 | 6.36 | 0.51 ± 0.01 | 0.0 |
| 4 | head $j = 1$ | 0.78 | 2.63 | 0.51 ± 0.01 | 0.28 |
| 5 | tail | 2.08 ± 0.08 | -0.39 | 0.51 ± 0.01 | 0.0 |
| 6 | tail | 1.59 ± 0.07 | -0.39 | 0.72 ± 0.01 | 0.0 |
| 7 | head $j = 0$ | 0.62 | 1.88 | 0.72 ± 0.01 | 0.0 |
| 8 | solvent | 0.00 ± 0.01 | 6.36 | 0.72 ± 0.01 | 0.0 |
| 9 | SiO$_2$ | 1.69 | 3.41 | 0.69 | 0.13 |
| 10 | Si | 0.00 | 2.07 | 0.39 | 0.0 |

Table S19: Sample 2, parameters at RH = 60%

| # | Slab | Thickness (nm) | SLD ($\times 10^{-4}$ nm$^{-2}$) | Roughness (nm) | vfsolv (%) |
|---|------|----------------|----------------------------------|----------------|------------|
| 0 | air | 0.00 | 0.00 | 0.00 | 0.0 |
| 1 | tail | 2.16 ± 0.08 | -0.39 | 0.49 ± 0.02 | 0.0 |
| 2 | head $j = 1$ | 0.81 | 2.63 | 0.49 ± 0.02 | 0.28 |
| 3 | solvent | 0.04 ± 0.04 | 6.36 | 0.49 ± 0.02 | 0.0 |
| 4 | head $j = 1$ | 0.81 | 2.63 | 0.49 ± 0.02 | 0.28 |
| 5 | tail | 2.16 ± 0.08 | -0.39 | 0.49 ± 0.02 | 0.0 |
| 6 | tail | 1.51 ± 0.07 | -0.39 | 0.68 ± 0.01 | 0.0 |
| 7 | head $j = 0$ | 0.59 | 1.88 | 0.68 ± 0.01 | 0.0 |
| 8 | solvent | 0.00 ± 0.01 | 6.36 | 0.68 ± 0.01 | 0.0 |
| 9 | SiO$_2$ | 1.69 | 3.41 | 0.69 | 0.13 |
| 10 | Si | 0.00 | 2.07 | 0.39 | 0.0 |

Table S20: Sample 2, parameters at RH = 70%

| # | Slab | Thickness (nm) | SLD ($\times 10^{-4}$ nm$^{-2}$) | Roughness (nm) | vfsolv (%) |
|---|---|---|---|---|---|
| 0 | air | 0.00 | 0.00 | 0.00 | 0.0 |
| 1 | tail | 2.18 ± 0.08 | -0.39 | 0.45 ± 0.02 | 0.0 |
| 2 | head $j = 1$ | 0.82 | 2.63 | 0.45 ± 0.02 | 0.28 |
| 3 | solvent | 0.05 ± 0.04 | 6.36 | 0.45 ± 0.02 | 0.0 |
| 4 | head $j = 1$ | 0.82 | 2.63 | 0.45 ± 0.02 | 0.28 |
| 5 | tail | 2.18 ± 0.08 | -0.39 | 0.45 ± 0.02 | 0.0 |
| 6 | tail | 1.50 ± 0.07 | -0.39 | 0.69 ± 0.01 | 0.0 |
| 7 | head $j = 0$ | 0.58 | 1.88 | 0.69 ± 0.01 | 0.0 |
| 8 | solvent | 0.00 ± 0.01 | 6.36 | 0.69 ± 0.01 | 0.0 |
| 9 | SiO$_2$ | 1.69 | 3.41 | 0.69 | 0.13 |
| 10 | Si | 0.00 | 2.07 | 0.39 | 0.0 |

Table S21: Sample 2, parameters at RH = 90%

| # | Slab | Thickness (nm) | SLD (×$10^{-4}$ nm$^{-2}$) | Roughness (nm) | vfsolv (%) |
|---|---|---|---|---|---|
| 0 | air | 0.00 | 0.00 | 0.00 | 0.0 |
| 1 | tail | 2.15 ± 0.07 | -0.39 | 0.47 ± 0.02 | 0.0 |
| 2 | head $j = 1$ | 0.81 | 2.63 | 0.47 ± 0.02 | 0.28 |
| 3 | solvent | 0.11 ± 0.04 | 6.36 | 0.47 ± 0.02 | 0.0 |
| 4 | head $j = 1$ | 0.81 | 2.63 | 0.47 ± 0.02 | 0.28 |
| 5 | tail | 2.15 ± 0.07 | -0.39 | 0.47 ± 0.02 | 0.0 |
| 6 | tail | 1.51 ± 0.06 | -0.39 | 0.65 ± 0.01 | 0.0 |
| 7 | head $j = 0$ | 0.59 | 1.88 | 0.65 ± 0.01 | 0.0 |
| 8 | solvent | 0.03 ± 0.01 | 6.36 | 0.65 ± 0.01 | 0.0 |
| 9 | SiO$_2$ | 1.69 | 3.41 | 0.69 | 0.13 |
| 10 | Si | 0.00 | 2.07 | 0.39 | 0.0 |

### 2.1.3 Sample 3: DPPS20-DPPS20-DPPS20 ($j = 0.2$)

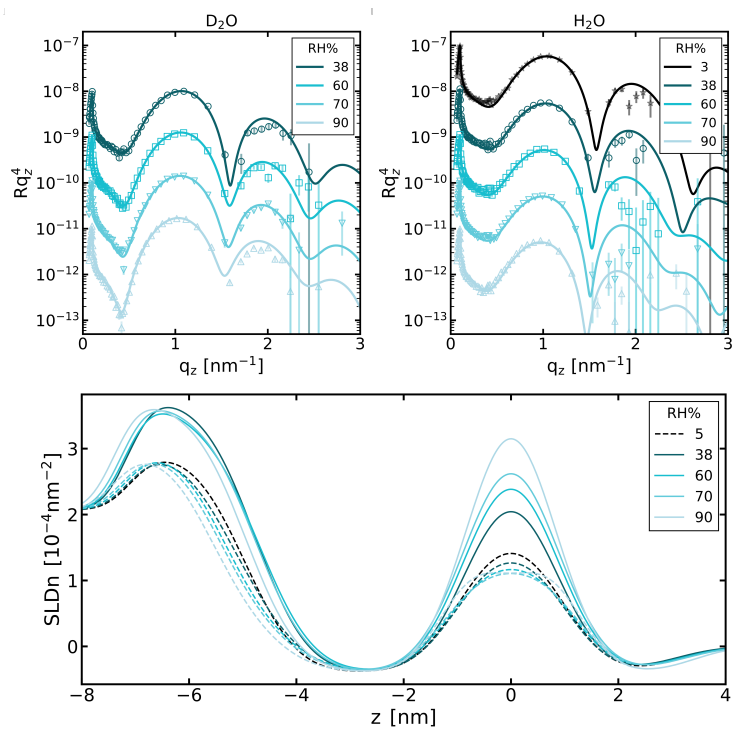

Figure S4: (Top) NR data of a DPPS20-DPPS20-DPPS20 trilayer (Sample 3) for different relative humidity. Data are presented as $Rq_z^4$ on a semilog scale with two contrasts: (Left) in $D_2O$ and (Right) in $H_2O$. Solid lines represent the best joined fits. (Bottom) $SLD_n$ profiles corresponding to the fits from above in $D_2O$ (Solid lines) and in $H_2O$ (Dashed line).

Table S22: Sample 3, parameters at RH = 3%

| # | Slab | Thickness (nm) | SLD ($\times 10^{-4}$ nm$^{-2}$) | Roughness (nm) | vfsolv (%) |
|---|---|---|---|---|---|
| 0 | air | 0.00 | 0.00 | 0.00 | 0.0 |
| 1 | tail | 2.22 ± 0.02 | -0.39 | 0.70 ± 0.04 | 0.0 |
| 2 | head $j$ = 0.2 | 0.86 | 2.03 | 0.70 ± 0.04 | 0.0 |
| 3 | solvent | 0.10 ± 0.04 | -0.56 | 0.70 ± 0.04 | 0.0 |
| 4 | head $j$ = 0.2 | 0.86 | 2.03 | 0.70 ± 0.04 | 0.0 |
| 5 | tail | 2.22 ± 0.02 | -0.39 | 0.70 ± 0.04 | 0.0 |
| 6 | tail | 1.64 | -0.39 | 0.72 ± 0.04 | 0.0 |
| 7 | head $j$ = 0.2 | 0.63 | 2.03 | 0.72 ± 0.04 | 0.0 |
| 8 | solvent | 0.00 ± 0.05 | -0.56 | 0.72 ± 0.04 | 0.0 |
| 9 | SiO$_2$ | 1.69 | 3.41 | 0.69 | 0.13 |
| 10 | Si | 0.00 | 2.07 | 0.39 | 0.0 |

Table S23: Sample 3, parameters at RH = 38%

| # | Slab | Thickness (nm) | SLD ($\times 10^{-4}$ nm$^{-2}$) | Roughness (nm) | vfsolv (%) |
|---|---|---|---|---|---|
| 0 | air | 0.00 | 0.00 | 0.00 | 0.00 |
| 1 | tail | 2.09 ± 0.01 | -0.39 | 0.70 ± 0.01 | 0.00 |
| 2 | head $j = 0.2$ | 0.81 | 2.03 | 0.70 ± 0.01 | 0.00 |
| 3 | solvent | 0.20 ± 0.00 | 6.36/-0.56 | 0.70 ± 0.01 | 0.00 |
| 4 | head $j = 0.2$ | 0.81 | 2.03 | 0.70 ± 0.01 | 0.00 |
| 5 | tail | 2.09 ± 0.01 | -0.39 | 0.70 ± 0.01 | 0.00 |
| 6 | tail | 1.64 ± 0.02 | -0.39 | 0.76 ± 0.01 | 0.00 |
| 7 | head $j = 0.2$ | 0.63 | 2.03 | 0.76 ± 0.01 | 0.00 |
| 8 | solvent | 0.17 ± 0.01 | 6.36 | 0.76 ± 0.01 | 0.00 |
| 9 | SiO$_2$ | 1.69 | 3.41 | 0.69 | 0.13 |
| 10 | Si | 0.00 | 2.07 | 0.39 | 0.00 |

Table S24: Sample 3, parameters at RH = 60%

| # | Slab | Thickness (nm) | SLD ($\times 10^{-4}$ nm$^{-2}$) | Roughness (nm) | vfsolv (%) |
|---|---|---|---|---|---|
| 0 | air | 0.00 | 0.00 | 0.00 | 0.00 |
| 1 | tail | 2.12 ± 0.01 | -0.39 | 0.71 ± 0.01 | 0.00 |
| 2 | head $j$ = 0.2 | 0.82 | 2.03 | 0.71 ± 0.01 | 0.00 |
| 3 | solvent | 0.32 ± 0.00 | 6.36/-0.56 | 0.71 ± 0.01 | 0.00 |
| 4 | head $j$ = 0.2 | 0.82 | 2.03 | 0.71 ± 0.01 | 0.00 |
| 5 | tail | 2.12 ± 0.01 | -0.39 | 0.71 ± 0.01 | 0.00 |
| 6 | tail | 1.59 ± 0.01 | -0.39 | 0.85 ± 0.01 | 0.00 |
| 7 | head $j$ = 0.2 | 0.61 | 2.03 | 0.85 ± 0.01 | 0.00 |
| 8 | solvent | 0.19 ± 0.01 | 6.36/-0.56 | 0.85 ± 0.01 | 0.00 |
| 9 | SiO$_2$ | 1.69 | 3.41 | 0.69 | 0.13 |
| 10 | Si | 0.00 | 2.07 | 0.39 | 0.00 |

Table S25: Sample 3, parameters at RH = 70%

| # | Slab | Thickness (nm) | SLD ($\times 10^{-4}$ nm$^{-2}$) | Roughness (nm) | vfsolv (%) |
|---|---|---|---|---|---|
| 0 | air | 0.00 | 0.00 | 0.00 | 0.00 |
| 1 | tail | 2.15 ± 0.01 | -0.39 | 0.71 ± 0.01 | 0.00 |
| 2 | head $j$ = 0.2 | 0.83 | 2.03 | 0.71 ± 0.01 | 0.00 |
| 3 | solvent | 0.39 ± 0.00 | 6.36/-0.56 | 0.71 ± 0.01 | 0.00 |
| 4 | head $j$ = 0.2 | 0.83 | 2.03 | 0.71 ± 0.01 | 0.00 |
| 5 | tail | 2.15 ± 0.01 | -0.39 | 0.71 ± 0.01 | 0.00 |
| 6 | tail | 1.57 ± 0.01 | -0.39 | 0.84 ± 0.01 | 0.00 |
| 7 | head $j$ = 0.2 | 0.60 | 2.03 | 0.84 ± 0.01 | 0.00 |
| 8 | solvent | 0.22 ± 0.01 | 6.36/-0.56 | 0.84 ± 0.01 | 0.00 |
| 9 | SiO$_2$ | 1.69 | 3.41 | 0.69 | 0.13 |
| 10 | Si | 0.00 | 2.07 | 0.39 | 0.00 |

Table S26: Sample 3, parameters at RH = 70%

| # | Slab | Thickness (nm) | SLD ($\times 10^{-4}$ nm$^{-2}$) | Roughness (nm) | vfsolv (%) |
|---|---|---|---|---|---|
| 0 | Air | 0.00 | 0.00 | 0.00 | 0.00 |
| 1 | tail | 2.30 ± 0.01 | -0.39 | 0.60 ± 0.01 | 0.00 |
| 2 | head $j$ = 0.2 | 0.89 | 2.03 | 0.60 ± 0.01 | 0.00 |
| 3 | solvent | 0.46 ± 0.00 | 6.36/-0.56 | 0.60 ± 0.01 | 0.00 |
| 4 | head $j$ = 0.2 | 0.89 | 2.03 | 0.60 ± 0.01 | 0.00 |
| 5 | tail | 2.30 ± 0.01 | -0.39 | 0.60 ± 0.01 | 0.00 |
| 6 | tail | 1.49 ± 0.01 | -0.39 | 0.82 ± 0.01 | 0.00 |
| 7 | head $j$ = 0.2 | 0.57 | 2.03 | 0.82 ± 0.01 | 0.00 |
| 8 | solvent | 0.22 ± 0.01 | 6.36/-0.56 | 0.82 ± 0.01 | 0.00 |
| 9 | SiO$_2$ | 1.69 | 3.41 | 0.69 | 0.13 |
| 10 | Si | 0.00 | 2.07 | 0.39 | 0.00 |

### 2.1.4 Sample 4: DPPC-DPPS60-DPPS60 ($j = 0.6$)

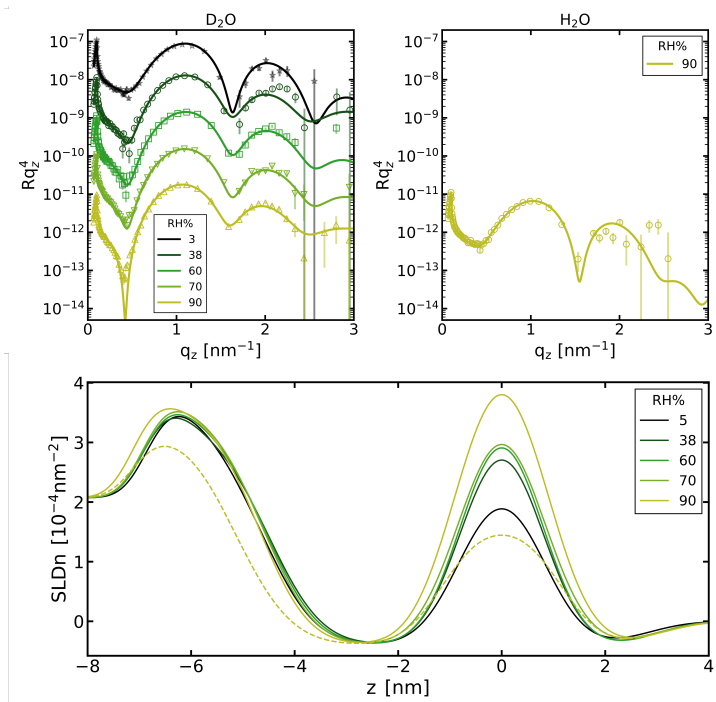

Figure S5: (Top) NR data of a DPPC-DPPS60-DPPS60 trilayer (Sample 4) for different relative humidity. Data are presented as $Rq_z^4$ on a semilog scale with two contrasts: (Left) in $D_2O$ and (Right) in $H_2O$. Solid lines represent the best joined fits. (Bottom) $SLD_n$ profiles corresponding to the fits from above in $D_2O$ (Solid lines) and in $H_2O$ (Dashed line).

Table S27: Sample 4, parameters at RH = 3%

| # | Slab | Thickness (nm) | SLD ($\times 10^{-4}$ nm$^{-2}$) | Roughness (nm) | vfsolv (%) |
|---|---|---|---|---|---|
| 0 | air | 0.00 | 0.00 | 0.00 | 0.00 |
| 1 | tail | 2.03 | -0.39 | 0.65 | 0.00 |
| 2 | head $j$ = 0.6 | 0.77 | 2.33 | 0.65 | 0.06 ± 0.10 |
| 3 | solvent | 0.00 ± 0.10 | 6.36 | 0.65 | 0.00 |
| 4 | head $j$ = 0.6 | 0.77 | 2.33 | 0.65 | 0.06 ± 0.10 |
| 5 | tail | 2.03 | -0.39 | 0.65 | 0.00 |
| 6 | tail | 1.69 | -0.39 | 0.79 | 0.00 |
| 7 | head $j$ = 0 | 0.66 | 1.88 | 0.79 | 0.00 ± 0.10 |
| 8 | solvent | 0.00 ± 0.04 | 6.36 | 0.79 | 0.00 |
| 9 | SiO$_2$ | 1.69 | 3.41 | 0.69 | 0.13 |
| 10 | Si | 0.00 | 2.07 | 0.39 | 0.00 |

Table S28: Sample 4, parameters at RH = 38%

| # | Slab | Thickness (nm) | SLD (×$10^{-4}$ nm$^{-2}$) | Roughness (nm) | vfsolv (%) |
|---|---|---|---|---|---|
| 0 | air | 0.00 | 0.00 | 0.00 | 0.00 |
| 1 | tail | 2.09 ± 0.12 | -0.39 | 0.60 ± 0.03 | 0.00 |
| 2 | head $j$ = 0.6 | 0.79 | 2.33 | 0.60 ± 0.03 | 0.00 |
| 3 | solvent | 0.26 ± 0.05 | 6.36 | 0.60 ± 0.03 | 0.00 |
| 4 | head $j$ = 0.6 | 0.79 | 2.33 | 0.60 ± 0.03 | 0.00 |
| 5 | tail | 2.09 ± 0.12 | -0.39 | 0.60 ± 0.03 | 0.00 |
| 6 | tail | 1.60 ± 0.12 | -0.39 | 0.86 ± 0.01 | 0.00 |
| 7 | head $j$ = 0 | 0.61 | 1.88 | 0.86 ± 0.01 | 0.25 |
| 8 | solvent | 0.00 ± 0.00 | 6.36 | 0.86 ± 0.01 | 0.00 |
| 9 | SiO$_2$ | 1.69 | 3.41 | 0.69 | 0.13 |
| 10 | Si | 0.00 | 2.07 | 0.39 | 0.00 |

Table S29: Sample 4, parameters at RH = 60%

| # | Slab | Thickness (nm) | SLD (×$10^{-4}$ nm$^{-2}$) | Roughness (nm) | vfsolv (%) |
|---|---|---|---|---|---|
| 0 | air | 0.00 | 0.00 | 0.00 | 0.00 |
| 1 | tail | 2.07 ± 0.08 | -0.39 | 0.59 ± 0.02 | 0.00 |
| 2 | head $j$ = 0.6 | 0.79 | 2.33 | 0.59 ± 0.02 | 0.00 |
| 3 | solvent | 0.32 ± 0.03 | 6.36 | 0.59 ± 0.02 | 0.00 |
| 4 | head $j$ = 0.6 | 0.79 | 2.33 | 0.59 ± 0.02 | 0.00 |
| 5 | tail | 2.07 ± 0.08 | -0.39 | 0.59 ± 0.02 | 0.00 |
| 6 | tail | 1.59 ± 0.08 | -0.39 | 0.82 ± 0.01 | 0.00 |
| 7 | head $j$ = 0 | 0.60 | 1.88 | 0.82 ± 0.01 | 0.25 |
| 8 | solvent | 0.00 ± 0.01 | 6.36 | 0.82 ± 0.01 | 0.00 |
| 9 | SiO$_2$ | 1.69 | 3.41 | 0.69 | 0.13 |
| 10 | Si | 0.00 | 2.07 | 0.39 | 0.00 |

Table S30: Sample 4, parameters at RH = 70%

| # | Slab | Thickness (nm) | SLD (×10$^{-4}$ nm$^{-2}$) | Roughness (nm) | vfsolv (%) |
|---|---|---|---|---|---|
| 0 | air | 0.00 | 0.00 | 0.00 | 0.00 |
| 1 | tail | 2.02 ± 0.08 | -0.39 | 0.65 ± 0.02 | 0.00 |
| 2 | head $j$ = 0.6 | 0.77 | 2.33 | 0.65 ± 0.02 | 0.00 |
| 3 | solvent | 0.41 ± 0.03 | 6.36 | 0.65 ± 0.02 | 0.00 |
| 4 | head $j$ = 0.6 | 0.77 | 2.33 | 0.65 ± 0.02 | 0.00 |
| 5 | tail | 2.02 ± 0.08 | -0.39 | 0.65 ± 0.02 | 0.00 |
| 6 | tail | 1.62 ± 0.08 | -0.39 | 0.78 ± 0.01 | 0.00 |
| 7 | head $j$ = 0 | 0.62 | 1.88 | 0.78 ± 0.01 | 0.25 |
| 8 | solvent | 0.00 ± 0.01 | 6.36 | 0.78 ± 0.01 | 0.00 |
| 9 | SiO$_2$ | 1.69 | 3.41 | 0.69 | 0.13 |
| 10 | Si | 0.00 | 2.07 | 0.39 | 0.00 |

Table S31: Sample 4, parameters at RH = 90%

| # | Slab | Thickness (nm) | SLD ($\times 10^{-4}$ nm$^{-2}$) | Roughness (nm) | vfsolv (%) |
|---|---|---|---|---|---|
| 0 | air | 0.00 | 0.00 | 0.00 | 0.0 |
| 1 | tail | 2.03 ± 0.04 | -0.39 | 0.65 ± 0.01 | 0.0 |
| 2 | head $j$ = 0.6 | 0.77 | 2.33 | 0.65 ± 0.01 | 0.21 ± 0.02 |
| 3 | solvent | 0.54 ± 0.03 | 6.36/-0.56 | 0.65 ± 0.01 | 0.0 |
| 4 | head $j$ = 0.6 | 0.77 | 2.33 | 0.65 ± 0.01 | 0.21 ± 0.02 |
| 5 | tail | 2.03 ± 0.04 | -0.39 | 0.65 ± 0.01 | 0.0 |
| 6 | tail | 1.69 ± 0.03 | -0.39 | 0.79 ± 0.01 | 0.0 |
| 7 | head $j$ = 0 | 0.66 | 1.88 | 0.79 ± 0.01 | 0.37 ± 0.04 |
| 8 | solvent | 0.01 ± 0.03 | 6.36/-0.56 | 0.79 ± 0.01 | 0.0 |
| 9 | SiO$_2$ | 1.69 | 3.41 | 0.69 | 13.0 |
| 10 | Si | 0.00 | 2.07 | 0.39 | 0.0 |

### 2.1.5 Sample 5: DPPC-DPPS70-DPPS70-DPPS70-DPPS70 ($j = 0.7$)

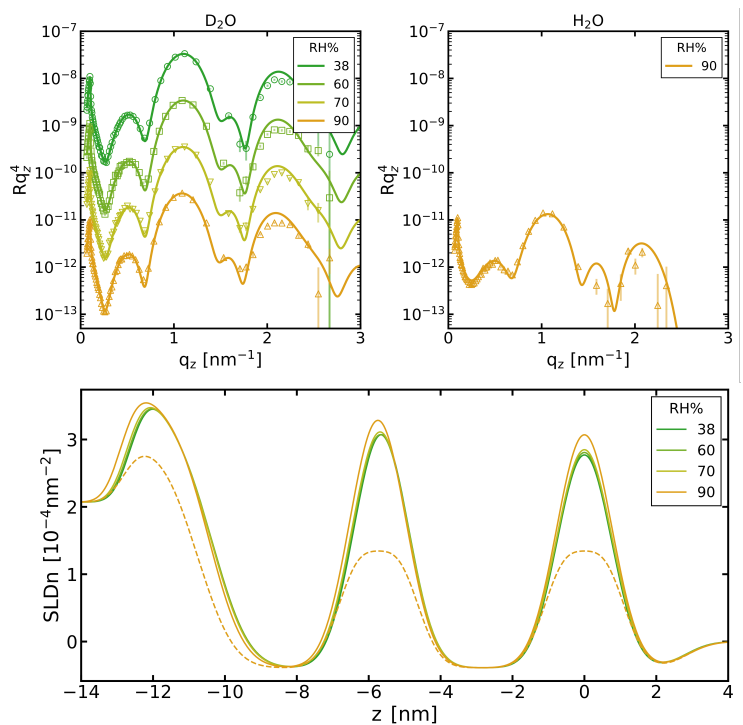

Figure S6: (Top) NR data of a DPPC-DPPS70-DPPS70-DPPS70-DPPS70 pentalayer (Sample 5) for different relative humidity. Data are presented as $Rq_z^4$ on a semilog scale with two contrasts: (Left) in $D_2O$ and (Right) in $H_2O$. Solid lines represent the best joined fits. (Bottom) $SLD_n$ profiles corresponding to the fits from above in $D_2O$ (Solid lines) and in $H_2O$ (Dashed line).

Table S32: Sample 5, parameters at RH = 38%

| #  | Slab         | Thickness (nm)  | SLD ($\times 10^{-4}$ nm$^{-2}$) | Roughness (nm)  | vfsolv (%)   |
|----|--------------|-----------------|----------------------------------|-----------------|--------------|
| 0  | air          | 0.00            | 0.00                             | 0.00            | 0.0          |
| 1  | tail         | 1.94            | -0.39                            | 0.56 ± 0.00     | 0.0          |
| 2  | head $j = 0.7$ | 0.73          | 2.40                             | 0.56 ± 0.00     | 0.0          |
| 3  | solvent      | 0.25 ± 0.00     | 6.36                             | 0.56 ± 0.00     | 0.0          |
| 4  | head $j = 0.7$ | 0.73          | 2.40                             | 0.56 ± 0.00     | 0.0          |
| 5  | tail         | 1.94            | -0.39                            | 0.56 ± 0.00     | 0.0          |
| 6  | tail         | 1.94            | -0.39                            | 0.56 ± 0.00     | 0.0          |
| 7  | head $j = 0.7$ | 0.73          | 2.40                             | 0.56 ± 0.00     | 0.0          |
| 8  | solvent      | 0.34 ± 0.00     | 6.36                             | 0.56 ± 0.00     | 0.0          |
| 9  | head $j = 0.7$ | 0.73          | 2.40                             | 0.56 ± 0.00     | 0.0          |
| 10 | tail         | 1.94            | -0.39                            | 0.56 ± 0.00     | 0.0          |
| 11 | tail         | 1.75 ± 0.12     | -0.39                            | 0.79 ± 0.02     | 0.0          |
| 12 | head $j = 0$ | 0.67            | 1.88                             | 0.79 ± 0.02     | 5.0 ± 3.4    |
| 13 | solvent      | 0.00 ± 0.17     | 6.36                             | 0.79 ± 0.02     | 0.0          |
| 14 | SiO$_2$      | 1.69            | 3.41                             | 0.69            | 13.0         |
| 15 | Si           | 0.00            | 2.07                             | 0.39            | 0.0          |

Table S33: Sample 5, parameters at RH = 60%

| # | Slab | Thickness (nm) | SLD (×$10^{-4}$ nm$^{-2}$) | Roughness (nm) | vfsolv (%) |
|---|------|---------------|---------------------------|----------------|------------|
| 0 | air | 0.00 | 0.00 | 0.00 | 0.0 |
| 1 | tail | 1.94 | -0.39 | 0.57 ± 0.00 | 0.0 |
| 2 | head $j$ = 0.7 | 0.73 | 2.40 | 0.57 ± 0.00 | 0.0 |
| 3 | solvent | 0.27 ± 0.00 | 6.36 | 0.57 ± 0.00 | 0.0 |
| 4 | head $j$ = 0.7 | 0.73 | 2.40 | 0.57 ± 0.00 | 0.0 |
| 5 | tail | 1.94 | -0.39 | 0.57 ± 0.00 | 0.0 |
| 6 | tail | 1.94 | -0.39 | 0.57 ± 0.00 | 0.0 |
| 7 | head $j$ = 0.7 | 0.73 | 2.40 | 0.57 ± 0.00 | 0.0 |
| 8 | solvent | 0.36 ± 0.00 | 6.36 | 0.57 ± 0.00 | 0.0 |
| 9 | head $j$ = 0.7 | 0.73 | 2.40 | 0.57 ± 0.00 | 0.0 |
| 10 | tail | 1.94 | -0.39 | 0.57 ± 0.00 | 0.0 |
| 11 | tail | 1.75 ± 0.12 | -0.39 | 0.79 ± 0.22 | 0.0 |
| 12 | head $j$ = 0 | 0.67 | 1.88 | 0.79 ± 0.22 | 9.0 ± 3.2 |
| 13 | solvent | 0.00 ± 0.16 | 6.36 | 0.79 ± 0.22 | 0.0 |
| 14 | SiO$_2$ | 1.69 | 3.41 | 0.69 | 13.0 |
| 15 | Si | 0.00 | 2.07 | 0.39 | 0.0 |

Table S34: Sample 5, parameters at RH = 70%

| # | Slab | Thickness (nm) | SLD ($\times 10^{-4}$ nm$^{-2}$) | Roughness (nm) | vfsolv (%) |
|---|---|---|---|---|---|
| 0 | air | 0.00 | 0.00 | 0.00 | 0.0 |
| 1 | tail | 1.94 | -0.39 | 0.57 ± 0.00 | 0.0 |
| 2 | head $j$ = 0.7 | 0.73 | 2.40 | 0.57 ± 0.00 | 0.0 |
| 3 | solvent | 0.28 ± 0.00 | 6.36 | 0.57 ± 0.00 | 0.0 |
| 4 | head $j$ = 0.7 | 0.73 | 2.40 | 0.57 ± 0.00 | 0.0 |
| 5 | tail | 1.94 | -0.39 | 0.57 ± 0.00 | 0.0 |
| 6 | tail | 1.94 | -0.39 | 0.57 ± 0.00 | 0.0 |
| 7 | head $j$ = 0.7 | 0.73 | 2.40 | 0.57 ± 0.00 | 0.0 |
| 8 | solvent | 0.37 ± 0.00 | 6.36 | 0.57 ± 0.00 | 0.0 |
| 9 | head $j$ = 0.7 | 0.73 | 2.40 | 0.57 ± 0.00 | 0.0 |
| 10 | tail | 1.94 | -0.39 | 0.57 ± 0.00 | 0.0 |
| 11 | tail | 1.77 ± 0.11 | -0.39 | 0.79 ± 0.02 | 0.0 |
| 12 | head $j$ = 0 | 0.68 | 1.88 | 0.79 ± 0.02 | 0.11 ± 0.29 |
| 13 | solvent | 0.01 ± 0.15 | 6.36 | 0.79 ± 0.02 | 0.0 |
| 14 | SiO$_2$ | 1.69 | 3.41 | 0.69 | 13.0 |
| 15 | Si | 0.00 | 2.07 | 0.39 | 0.0 |

Table S35: Sample 5, parameters at RH = 90%

| # | Slab | Thickness (nm) | SLD (×$10^{-4}$ nm$^{-2}$) | Roughness (nm) | vfsolv (%) |
|---|---|---|---|---|---|
| 0 | air | 0.00 | 0.00 | 0.00 | 0.0 |
| 1 | tail | 1.94 ± 0.10 | -0.39 | 0.57 ± 0.00 | 0.0 |
| 2 | head $j$ = 0.7 | 0.73 | 2.40 | 0.57 ± 0.00 | 0.0 |
| 3 | solvent | 0.35 ± 0.00 | 6.36/-0.56 | 0.57 ± 0.00 | 0.0 |
| 4 | head $j$ = 0.7 | 0.73 | 2.40 | 0.57 ± 0.00 | 0.0 |
| 5 | tail | 1.94 ± 0.10 | -0.39 | 0.57 ± 0.00 | 0.0 |
| 6 | tail | 1.94 ± 0.10 | -0.39 | 0.57 ± 0.00 | 0.0 |
| 7 | head $j$ = 0.7 | 0.73 | 2.40 | 0.57 ± 0.00 | 0.0 |
| 8 | solvent | 0.42 ± 0.00 | 6.36/-0.56 | 0.57 ± 0.00 | 0.0 |
| 9 | head $j$ = 0.7 | 0.73 | 2.40 | 0.57 ± 0.00 | 0.0 |
| 10 | tail | 1.94 ± 0.10 | -0.39 | 0.57 ± 0.00 | 0.0 |
| 11 | tail | 1.84 ± 0.10 | -0.39 | 0.78 ± 0.00 | 0.0 |
| 12 | head $j$ = 0 | 0.71 | 1.88 | 0.78 ± 0.00 | 0.28 ± 0.25 |
| 13 | solvent | 0.00 ± 0.13 | 6.36/-0.56 | 0.78 ± 0.00 | 0.0 |
| 14 | SiO$_2$ | 1.69 | 3.41 | 0.69 | 13.0 |
| 15 | Si | 0.00 | 2.07 | 0.39 | 0.0 |

### 2.1.6 Sample 6: DPPC-DPPS80-DPPS80-DPPS80-DPPS80 ($j = 0.8$)

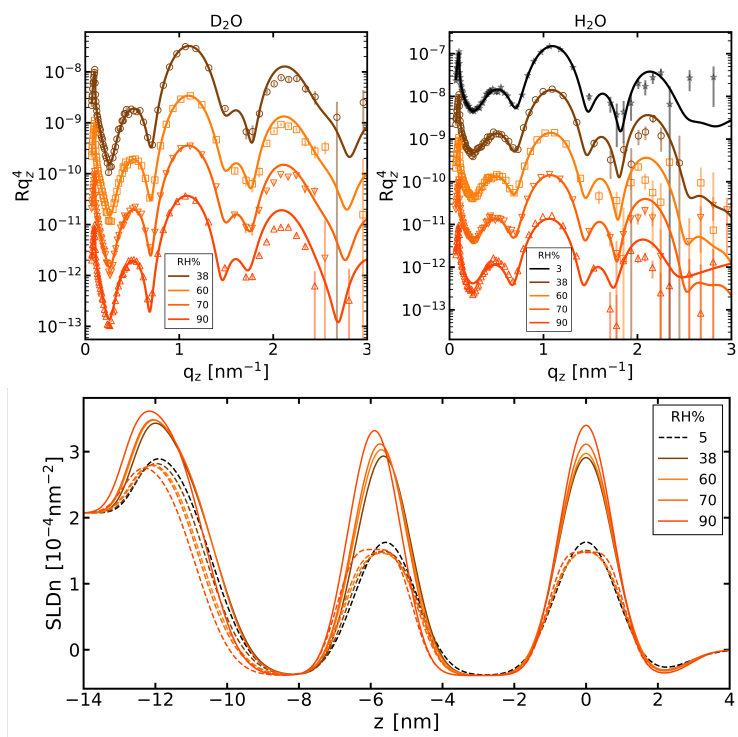

Figure S7: (Top) NR data of a DPPC-DPPS80-DPPS80-DPPS80-DPPS80 pentalayer (Sample 6) for different relative humidity. Data are presented as $Rq_z^4$ on a semilog scale with two contrasts: (Left) in $D_2O$ and (Right) in $H_2O$. Solid lines represent the best joined fits. (Bottom) $SLD_n$ profiles corresponding to the fits from above in $D_2O$ (Solid lines) and in $H_2O$ (Dashed line).

Table S36: Sample 6, parameters at RH = 3%

| # | Slab | Thickness (nm) | SLD (×$10^{-4}$ nm$^{-2}$) | Roughness (nm) | vfsolv (%) |
|---|---|---|---|---|---|
| 0 | air | 0.00 | 0.00 | 0.00 | 0.0 |
| 1 | tail | 2.00 ± 0.02 | -0.39 | 0.68 ± 0.02 | 0.0 |
| 2 | head $j$ = 0.8 | 0.74 ± 0.01 | 2.55 | 0.68 ± 0.02 | 0.05 ± 0.02 |
| 3 | solvent | 0.11 ± 0.06 | -0.56 | 0.68 ± 0.02 | 0.0 |
| 4 | head $j$ = 0.8 | 0.74 ± 0.01 | 2.55 | 0.68 ± 0.02 | 0.05 ± 0.02 |
| 5 | tail | 2.00 ± 0.02 | -0.39 | 0.68 ± 0.02 | 0.0 |
| 6 | tail | 2.00 ± 0.02 | -0.39 | 0.68 ± 0.02 | 0.0 |
| 7 | head $j$ = 0.8 | 0.74 ± 0.01 | 2.55 | 0.68 ± 0.02 | 0.05 ± 0.02 |
| 8 | solvent | 0.12 ± 0.06 | -0.56 | 0.68 ± 0.02 | 0.0 |
| 9 | head $j$ = 0.8 | 0.74 ± 0.01 | 2.55 | 0.68 ± 0.02 | 0.05 ± 0.02 |
| 10 | tail | 2.00 ± 0.02 | -0.39 | 0.68 ± 0.02 | 0.0 |
| 11 | tail | 1.72 ± 0.02 | -0.39 | 0.75 ± 0.08 | 0.0 |
| 12 | head $j$ = 0 | 0.67 ± 0.01 | 1.88 | 0.75 ± 0.08 | 0.07 ± 0.09 |
| 13 | solvent | 0.07 ± 0.05 | -0.56 | 0.75 ± 0.08 | 0.0 |
| 14 | SiO$_2$ | 1.69 | 3.41 | 0.69 | 13.0 |
| 15 | Si | 0.00 | 2.07 | 0.39 | 0.0 |

Table S37: Sample 6, parameters at RH = 38%

| # | Slab | Thickness (nm) | SLD (×10⁻⁴ nm⁻²) | Roughness (nm) | vfsolv (%) |
|---|---|---|---|---|---|
| 0 | air | 0.00 | 0.00 | 0.00 | 0.0 |
| 1 | tail | 1.94 ± 0.00 | -0.39 | 0.57 ± 0.01 | 0.0 |
| 2 | head $j$ = 0.8 | 0.73 | 2.48 | 0.57 ± 0.01 | 0.0 ± 0.01 |
| 3 | solvent | 0.29 ± 0.01 | 6.36/-0.56 | 0.57 ± 0.01 | 0.0 |
| 4 | head $j$ = 0.8 | 0.73 | 2.48 | 0.57 ± 0.01 | 0.0 ± 0.01 |
| 5 | tail | 1.94 ± 0.00 | -0.39 | 0.57 ± 0.01 | 0.0 |
| 6 | tail | 1.94 ± 0.00 | -0.39 | 0.57 ± 0.01 | 0.0 |
| 7 | head $j$ = 0.8 | 0.73 | 2.48 | 0.57 ± 0.01 | 0.0 ± 0.01 |
| 8 | solvent | 0.30 ± 0.01 | 6.36/-0.56 | 0.57 ± 0.01 | 0.0 |
| 9 | head $j$ = 0.8 | 0.73 | 2.48 | 0.57 ± 0.01 | 0.0 ± 0.01 |
| 10 | tail | 1.94 ± 0.00 | -0.39 | 0.57 ± 0.01 | 0.0 |
| 11 | tail | 1.75 ± 0.02 | -0.39 | 0.80 ± 0.01 | 0.0 |
| 12 | head $j$ = 0 | 0.68 | 1.88 | 0.80 ± 0.01 | 0.03 ± 0.05 |
| 13 | solvent | 0.01 ± 0.03 | 6.36/-0.56 | 0.80 ± 0.01 | 0.0 |
| 14 | SiO$_2$ | 1.69 | 3.41 | 0.69 | 13.0 |
| 15 | Si | 0.00 | 2.07 | 0.39 | 0.0 |

Table S38: Sample 6, parameters at RH = 60%

| # | Slab | Thickness (nm) | SLD (×$10^{-4}$ nm$^{-2}$) | Roughness (nm) | vfsolv (%) |
|---|---|---|---|---|---|
| 0 | air | 0.00 | 0.00 | 0.00 | 0.0 |
| 1 | tail | 1.95 ± 0.00 | -0.39 | 0.57 ± 0.01 | 0.0 |
| 2 | head $j$ = 0.8 | 0.74 | 2.48 | 0.57 ± 0.01 | 0.0 ± 0.0 |
| 3 | solvent | 0.31 ± 0.01 | 6.36/-0.56 | 0.57 ± 0.01 | 0.0 |
| 4 | head $j$ = 0.8 | 0.74 | 2.48 | 0.57 ± 0.01 | 0.0 ± 0.0 |
| 5 | tail | 1.95 ± 0.00 | -0.39 | 0.57 ± 0.01 | 0.0 |
| 6 | tail | 1.95 ± 0.00 | -0.39 | 0.57 ± 0.01 | 0.0 |
| 7 | head $j$ = 0.8 | 0.74 | 2.48 | 0.57 ± 0.01 | 0.0 ± 0.0 |
| 8 | solvent | 0.33 ± 0.01 | 6.36/-0.56 | 0.57 ± 0.01 | 0.0 |
| 9 | head $j$ = 0.8 | 0.74 | 2.48 | 0.57 ± 0.01 | 0.0 ± 0.0 |
| 10 | tail | 1.95 ± 0.00 | -0.39 | 0.57 ± 0.01 | 0.0 |
| 11 | tail | 1.75 ± 0.01 | -0.39 | 0.79 ± 0.01 | 0.0 |
| 12 | head $j$ = 0 | 0.68 | 1.88 | 0.79 ± 0.01 | 0.14 ± 0.04 |
| 13 | solvent | 0.00 ± 0.03 | 6.36/-0.56 | 0.79 ± 0.01 | 0.0 |
| 14 | SiO$_2$ | 1.69 | 3.41 | 0.69 | 13.0 |
| 15 | Si | 0.00 | 2.07 | 0.39 | 0.0 |

Table S39: Sample 6, parameters at RH = 70%

| # | Slab | Thickness (nm) | SLD ($\times 10^{-4}$ nm$^{-2}$) | Roughness (nm) | vfsolv (%) |
|---|---|---|---|---|---|
| 0 | air | 0.00 | 0.00 | 0.00 | 0.0 |
| 1 | tail | 1.96 ± 0.00 | -0.39 | 0.55 ± 0.01 | 0.0 |
| 2 | head $j$ = 0.8 | 0.74 | 2.48 | 0.55 ± 0.01 | 0.0 ± 0.0 |
| 3 | solvent | 0.33 ± 0.01 | 6.36/-0.56 | 0.55 ± 0.01 | 0.0 |
| 4 | head $j$ = 0.8 | 0.74 | 2.48 | 0.55 ± 0.01 | 0.0 ± 0.0 |
| 5 | tail | 1.96 ± 0.00 | -0.39 | 0.55 ± 0.01 | 0.0 |
| 6 | tail | 1.96 ± 0.00 | -0.39 | 0.55 ± 0.01 | 0.0 |
| 7 | head $j$ = 0.8 | 0.74 | 2.48 | 0.55 ± 0.01 | 0.0 ± 0.0 |
| 8 | solvent | 0.33 ± 0.01 | 6.36/-0.56 | 0.55 ± 0.01 | 0.0 |
| 9 | head $j$ = 0.8 | 0.74 | 2.48 | 0.55 ± 0.01 | 0.0 ± 0.0 |
| 10 | tail | 1.96 ± 0.00 | -0.39 | 0.55 ± 0.01 | 0.0 |
| 11 | tail | 1.73 ± 0.01 | -0.39 | 0.80 ± 0.01 | 0.0 |
| 12 | head $j$ = 0 | 0.67 | 1.88 | 0.80 ± 0.01 | 0.18 ± 0.04 |
| 13 | solvent | 0.01 ± 0.03 | 6.36/-0.56 | 0.80 ± 0.01 | 0.0 |
| 14 | SiO$_2$ | 1.69 | 3.41 | 0.69 | 13.0 |
| 15 | Si | 0.00 | 2.07 | 0.39 | 0.0 |

Table S40: Sample 6, parameters at RH = 90%

| # | Slab | Thickness (nm) | SLD ($\times 10^{-4}$ nm$^{-2}$) | Roughness (nm) | vfsolv (%) |
|---|---|---|---|---|---|
| 0 | air | 0.00 | 0.00 | 0.00 | 0.0 |
| 1 | tail | 2.02 ± 0.00 | -0.39 | 0.49 ± 0.01 | 0.0 |
| 2 | head $j$ = 0.8 | 0.76 | 2.48 | 0.49 ± 0.01 | 0.0 ± 0.0 |
| 3 | solvent | 0.35 ± 0.01 | 6.36/-0.56 | 0.49 ± 0.01 | 0.0 |
| 4 | head $j$ = 0.8 | 0.76 | 2.48 | 0.49 ± 0.01 | 0.0 ± 0.0 |
| 5 | tail | 2.02 ± 0.00 | -0.39 | 0.49 ± 0.01 | 0.0 |
| 6 | tail | 2.02 ± 0.00 | -0.39 | 0.49 ± 0.01 | 0.0 |
| 7 | head $j$ = 0.8 | 0.76 | 2.48 | 0.49 ± 0.01 | 0.0 ± 0.0 |
| 8 | solvent | 0.33 ± 0.01 | 6.36/-0.56 | 0.49 ± 0.01 | 0.0 |
| 9 | head $j$ = 0.8 | 0.76 | 2.48 | 0.49 ± 0.01 | 0.0 ± 0.0 |
| 10 | tail | 2.02 ± 0.00 | -0.39 | 0.49 ± 0.01 | 0.0 |
| 11 | tail | 1.70 ± 0.01 | -0.39 | 0.71 ± 0.01 | 0.0 |
| 12 | head $j$ = 0 | 0.66 | 1.88 | 0.71 ± 0.01 | 0.23 ± 0.03 |
| 13 | solvent | 0.01 ± 0.02 | 6.36/-0.56 | 0.71 ± 0.01 | 0.0 |
| 14 | SiO$_2$ | 1.69 | 3.41 | 0.69 | 13.0 |
| 15 | Si | 0.00 | 2.07 | 0.39 | 0.0 |

## 2.2 X-ray Reflectivity

### 2.2.1 Sample 7: DPPC-DPPC-DPPC ($j = 0$)

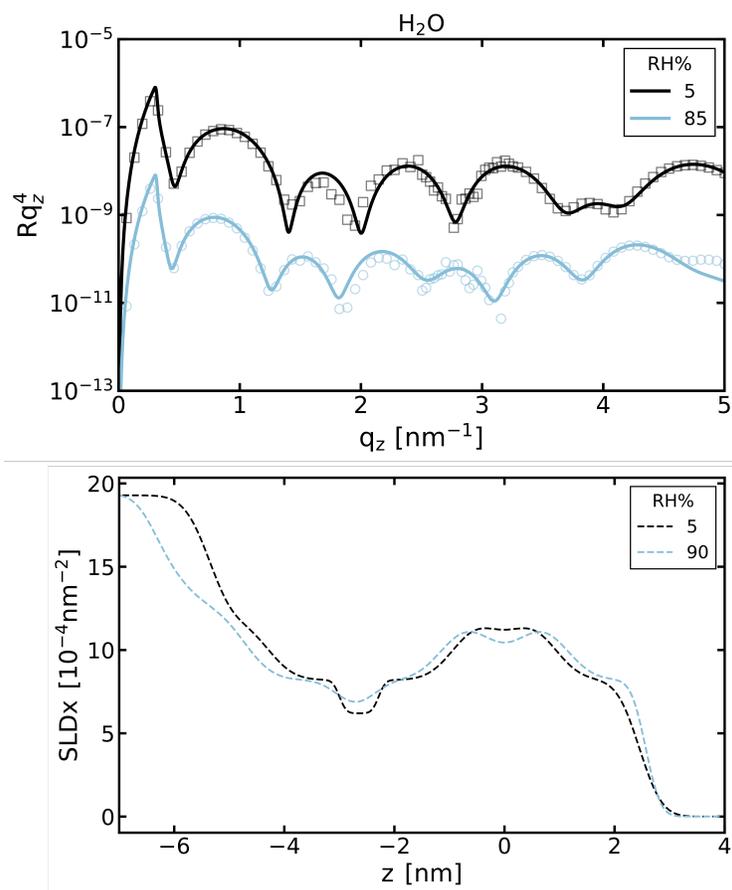

Figure S8: (Top) XRR data of a pure DPPC-DPPC-DPPC trilayer (Sample 7) in $H_2O$ for 2 different relative humidity. Data are presented as $Rq_z^4$ on a semilog scale. Solid lines represent the best fits. (Bottom) $SLD_x$ profiles corresponding to the fits from above.

Table S41: Sample 7, parameters at RH = 5%

| # | Slab | Thickness (nm) | SLD ($\times 10^{-4}$ nm$^{-2}$) | Roughness (nm) | vfsolv (%) |
|---|---|---|---|---|---|
| 0 | air | 0.00 | 0.00 | 0.00 | 0.0 |
| 1 | tail | 1.64 ± 0.02 | 8.2 | 0.32 ± 0.00 | 0.0 |
| 2 | head $j = 0$ | 0.55 | 14.45 | 0.41 ± 0.01 | 0.22 |
| 3 | H$_2$O | 0.53 ± 0.01 | 9.41 | 0.41 ± 0.01 | 0.0 |
| 4 | head $j = 0$ | 0.55 | 14.45 | 0.41 ± 0.01 | 0.22 |
| 5 | tail | 1.47 ± 0.01 | 8.2 | 0.41 ± 0.01 | 0.0 |
| 6 | tail CH$_3$ | 0.71 ± 0.02 | 6.2 | 0.10 ± 0.01 | 0.0 |
| 7 | tail | 1.40 ± 0.02 | 8.2 | 0.10 ± 0.01 | 0.0 |
| 8 | head $j = 0$ | 0.53 | 14.45 | 0.41 ± 0.01 | 0.22 |
| 9 | H$_2$O | 0.35 ± 0.15 | 9.41 | 0.41 ± 0.01 | 0.0 |
| 10 | SiO$_2$ | 0.49 ± 0.39 | 18.5 ± 0.39 | 0.35 ± 0.07 | 0.0 |
| 11 | Si | 0.00 | 19.29 ± 0.00 | 0.30 | 0.0 |

Table S42: Sample 7, parameters at RH = 85%

| # | Slab | Thickness (nm) | SLD ($\times 10^{-4}$ nm$^{-2}$) | Roughness (nm) | vfsolv (%) |
|---|---|---|---|---|---|
| 0 | air | 0.00 | 0.00 | 0.00 | 0.0 |
| 1 | tail | 1.55 ± 0.01 | 8.2 | 0.21 ± 0.00 | 0.0 |
| 2 | head $j = 0$ | 0.56 | 14.45 | 0.43 ± 0.00 | 0.22 |
| 3 | H$_2$O | 0.91 ± 0.01 | 9.41 | 0.43 ± 0.00 | 0.0 |
| 4 | head $j = 0$ | 0.56 | 14.45 | 0.43 ± 0.00 | 0.22 |
| 5 | tail | 1.40 ± 0.01 | 8.2 | 0.43 ± 0.00 | 0.0 |
| 6 | tail CH$_3$ | 0.58 ± 0.02 | 6.2 | 0.30 ± 0.01 | 0.0 |
| 7 | tail | 1.78 ± 0.01 | 8.2 | 0.30 ± 0.01 | 0.0 |
| 8 | head $j = 0$ | 0.71 | 14.45 | 0.43 ± 0.00 | 0.22 |
| 9 | H$_2$O | 0.22 ± 0.03 | 9.41 | 0.43 ± 0.00 | 0.0 |
| 10 | SiO$_2$ | 0.68 ± 0.02 | 15.5 ± 0.03 | 0.42 ± 0.02 | 0.0 |
| 11 | Si | 0.00 | 19.37 ± 0.00 | 0.30 | 0.0 |

### 2.2.2 Sample 8: DSPC-DPPS-DPPS ($j = 1$)

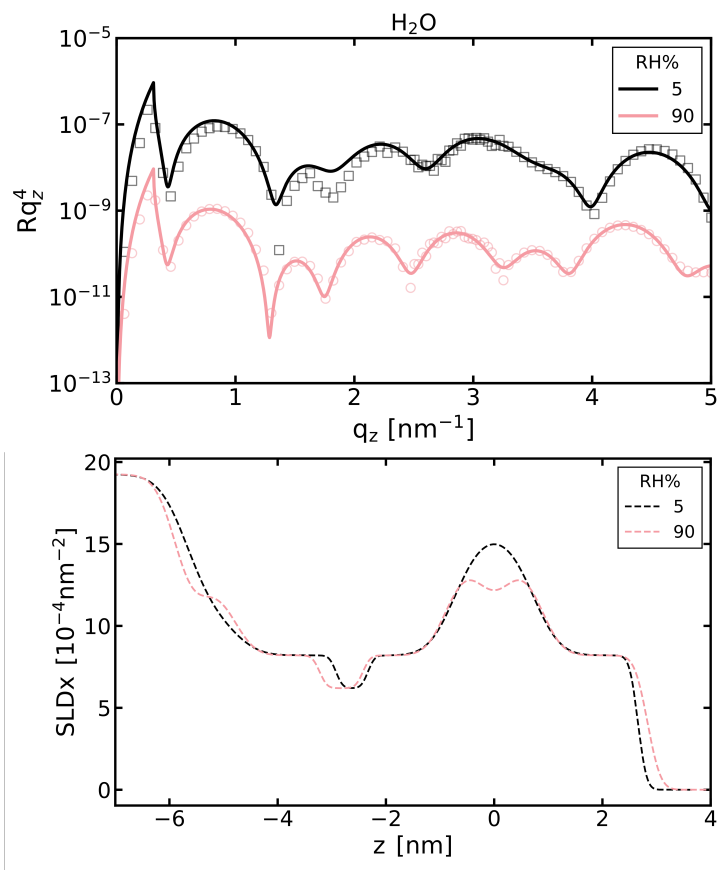

Figure S9: (Top) XRR data of a DSPC-DPPS-DPPS trilayer (Sample 8) in $H_2O$ for 2 different relative humidity. Data are presented as $Rq_z^4$ on a semilog scale. Solid lines represent the best fits. (Bottom) $SLD_x$ profiles corresponding to the fits from above.

Table S43: Sample 8, parameters at RH = 5%

| # | Slab | Thickness (nm) | SLD (×$10^{-4}$ nm$^{-2}$) | Roughness (nm) | vfsolv (%) |
|---|---|---|---|---|---|
| 0 | air | 0.00 | 0.00 | 0.00 | 0.0 |
| 1 | tail | 1.94 ± 0.00 | 8.2 | 0.12 ± 0.00 | 0.0 |
| 2 | head $j = 1$ | 0.72 | 15.5 | 0.39 ± 0.00 | 0.0 ± 0.0 |
| 3 | H$_2$O | 0.00 ± 0.00 | 9.41 | 0.39 ± 0.00 | 0.0 |
| 4 | head $j = 1$ | 0.72 | 15.5 | 0.39 ± 0.00 | 0.0 ± 0.0 |
| 5 | tail | 1.61 ± 0.01 | 8.2 | 0.39 ± 0.00 | 0.0 |
| 6 | tail CH$_3$ | 0.59 ± 0.02 | 6.2 | 0.10 ± 0.01 | 0.0 |
| 7 | tail | 1.89 ± 0.03 | 8.2 | 0.10 ± 0.01 | 0.0 |
| 8 | head $j = 0$ | 0.76 | 14.45 | 0.39 ± 0.00 | 0.07 ± 0.01 |
| 9 | H$_2$O | 0.09 ± 0.02 | 9.41 | 0.39 ± 0.00 | 0.0 |
| 10 | SiO$_2$ | 0.66 ± 0.06 | 19 | 0.40 ± 0.01 | 0.0 |
| 11 | Si | 0.00 | 19.24 | 0.30 | 0.0 |

Table S44: Sample 8, parameters at RH = 90%

| # | Slab | Thickness (nm) | SLD ($\times 10^{-4}$ nm$^{-2}$) | Roughness (nm) | vfsolv (%) |
|---|---|---|---|---|---|
| 0 | air | 0.00 | 0.00 | 0.00 | 0.0 |
| 1 | tail | 1.96 ± 0.00 | 8.2 | 0.19 ± 0.00 | 0.0 |
| 2 | head $j = 1$ | 0.73 | 15.5 | 0.30 ± 0.00 | 0.28 ± 0.01 |
| 3 | H$_2$O | 0.29 ± 0.00 | 9.41 | 0.30 ± 0.00 | 0.0 |
| 4 | head $j = 1$ | 0.73 | 15.5 | 0.30 ± 0.00 | 0.28 ± 0.01 |
| 5 | tail | 1.55 ± 0.01 | 8.2 | 0.30 ± 0.00 | 0.0 |
| 6 | tail CH$_3$ | 0.80 ± 0.01 | 6.2 | 0.11 ± 0.01 | 0.0 |
| 7 | tail | 1.58 ± 0.01 | 8.2 | 0.11 ± 0.01 | 0.0 |
| 8 | head $j = 0$ | 0.69 | 14.5 | 0.30 ± 0.00 | 0.39 ± 0.02 |
| 9 | H$_2$O | 0.36 ± 0.07 | 9.41 | 0.30 ± 0.00 | 0.0 |
| 10 | SiO$_2$ | 0.49 ± 1.13 | 19.0 ± 0.18 | 0.30 ± 0.04 | 0.0 |
| 11 | Si | 0.00 | 19.25 ± 0.00 | 0.30 | 0.0 |

### 2.2.3 Sample 9: DSPC-DPPS40-DPPS40 ($j = 0.4$)

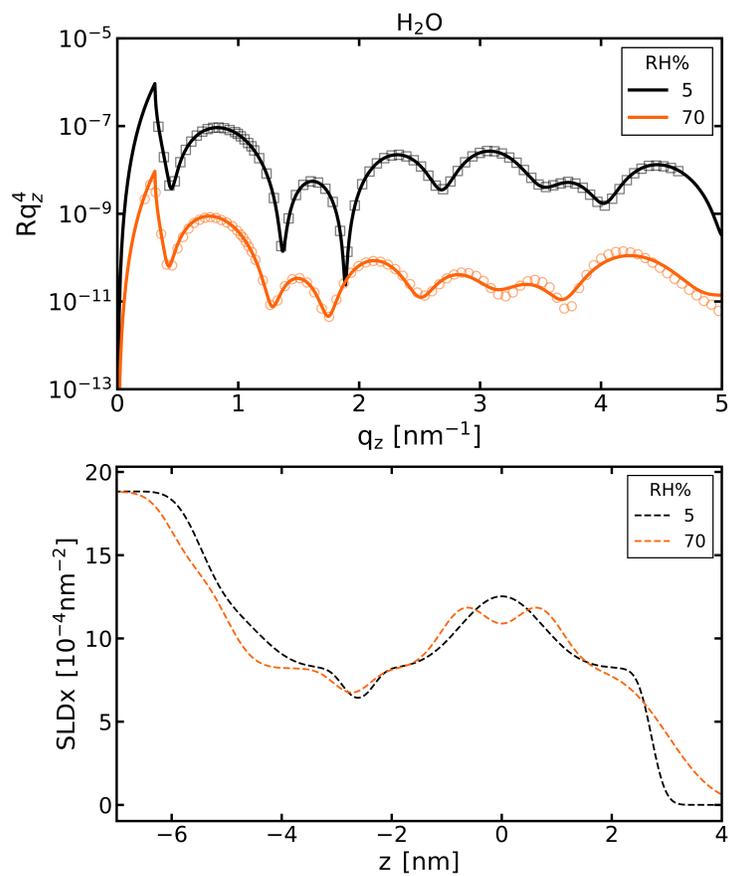

Figure S10: (Top) XRR data of a DSPC-DPPS40-DPPS40 trilayer (Sample 9) in $H_2O$ for 2 different relative humidity. Data are presented as $Rq_z^4$ on a semilog scale. Solid lines represent the best fits. (Bottom) $SLD_x$ profiles corresponding to the fits from above.

Table S45: Sample 9, parameters at RH = 5%

| # | Slab | Thickness (nm) | SLD ($\times 10^{-4}$ nm$^{-2}$) | Roughness (nm) | vfsolv (%) |
|---|---|---|---|---|---|
| 0 | air | 0.00 | 0.00 | 0.00 | 0.0 |
| 1 | tail | 2.00 ± 0.00 | 8.2 | 0.20 ± 0.00 | 0.0 |
| 2 | head $j$ = 0.4 | 0.72 | 14.87 | 0.56 ± 0.00 | 0.24 ± 0.0 |
| 3 | H$_2$O | 0.00 ± 0.00 | 9.41 | 0.56 ± 0.00 | 0.0 |
| 4 | head $j$ = 0.4 | 0.72 | 14.87 | 0.56 ± 0.00 | 0.24 ± 0.0 |
| 5 | tail | 1.60 ± 0.00 | 8.2 | 0.56 ± 0.00 | 0.0 |
| 6 | tail CH$_3$ | 0.58 ± 0.01 | 6.2 | 0.18 ± 0.00 | 0.0 |
| 7 | tail | 1.66 ± 0.00 | 8.2 | 0.18 ± 0.00 | 0.0 |
| 8 | head $j$ = 0 | 0.68 | 14.5 | 0.56 ± 0.00 | 0.07 ± 0.01 |
| 9 | H$_2$O | 0.23 ± 0.00 | 9.41 | 0.56 ± 0.00 | 0.0 |
| 10 | SiO$_2$ | 2.04 ± 0.00 | 18.82 | 0.40 ± 0.00 | 0.0 |
| 11 | Si | 0.00 | 19.35 | 0.01 | 0.0 |

Table S46: Sample 9, parameters at RH = 75%

| # | Slab | Thickness (nm) | SLD ($\times 10^{-4}$ nm$^{-2}$) | Roughness (nm) | vfsolv (%) |
|---|---|---|---|---|---|
| 0 | air | 0.00 | 0.00 | 0.00 | 0.0 |
| 1 | tail | 2.02 ± 0.00 | 8.2 | 0.67 ± 0.00 | 0.0 |
| 2 | head $j$ = 0.4 | 0.65 | 14.87 | 0.39 ± 0.00 | 0.20 ± 0.0 |
| 3 | H$_2$O | 0.73 ± 0.00 | 9.41 | 0.39 ± 0.00 | 0.0 |
| 4 | head $j$ = 0.4 | 0.65 | 14.87 | 0.39 ± 0.00 | 0.20 ± 0.0 |
| 5 | tail | 1.41 ± 0.00 | 8.2 | 0.39 ± 0.00 | 0.0 |
| 6 | tail CH$_3$ | 0.59 ± 0.01 | 6.2 | 0.27 ± 0.00 | 0.0 |
| 7 | tail | 2.00 ± 0.00 | 8.2 | 0.27 ± 0.00 | 0.0 |
| 8 | head $j$ = 0 | 0.80 | 14.5 | 0.39 ± 0.00 | 0.00 ± 0.01 |
| 9 | H$_2$O | 0.11 ± 0.00 | 9.41 | 0.39 ± 0.00 | 0.0 |
| 10 | SiO$_2$ | 1.91 ± 0.00 | 18.82 | 0.36 ± 0.00 | 0.0 |
| 11 | Si | 0.00 | 19.35 | 0.01 | 0.0 |

### 2.2.4 Sample 10: DSPC-DPPS50-DPPS50 ($j = 0.5$)- CsOH 0.1 mM

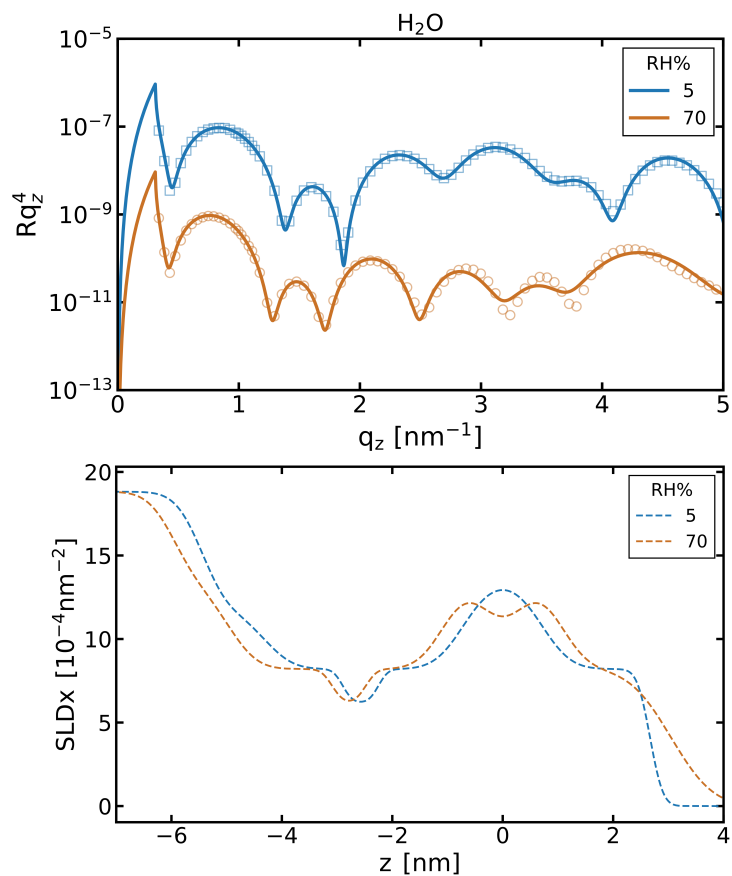

Figure S11: (Top) XRR data of a DSPC-DPPS50-DPPS50 trilayer (Sample 10) in $H_2O$ (CsOH 0.1 mM solution) for 2 different relative humidity. Data are presented as $Rq_z^4$ on a semilog scale. Solid lines represent the best fits. (Bottom) $SLD_x$ profiles corresponding to the fits from above.

Table S47: Sample 10, parameters at RH = 5%

| # | Slab | Thickness (nm) | SLD (×$10^{-4}$ nm$^{-2}$) | Roughness (nm) | vfsolv (%) |
|---|---|---|---|---|---|
| 0 | air | 0.00 | 0.00 | 0.00 | 0.0 |
| 1 | tail | 1.96 ± 0.00 | 8.2 | 0.18 ± 0.00 | 0.0 |
| 2 | head $j$ = 0.5 | 0.71 | 14.98 | 0.42 ± 0.00 | 0.28 |
| 3 | $H_2O$ | 0.00 ± 0.00 | 9.41 | 0.42 ± 0.00 | 0.0 |
| 4 | head mix | 0.71 | 14.98 | 0.42 ± 0.00 | 0.28 |
| 5 | tail | 1.54 ± 0.00 | 8.2 | 0.42 ± 0.00 | 0.0 |
| 6 | tail $CH_3$ | 0.65 ± 0.01 | 6.2 | 0.14 ± 0.00 | 0.0 |
| 7 | tail | 1.47 ± 0.00 | 8.2 | 0.14 ± 0.00 | 0.0 |
| 8 | head $j$ = 0 | 0.63 | 14.45 | 0.42 ± 0.00 | 0.41 |
| 9 | $H_2O$ | 0.39 ± 0.00 | 9.41 | 0.42 ± 0.00 | 0.0 |
| 10 | $SiO_2$ | 1.85 ± 0.00 | 18.82 | 0.41 ± 0.00 | 0.0 |
| 11 | Si | 0.00 | 19.35 | 0.01 | 0.0 |

Table S48: Sample 10, parameters at RH = 70%

| # | Slab | Thickness (nm) | SLD (×$10^{-4}$ nm$^{-2}$) | Roughness (nm) | vfsolv (%) |
|---|---|---|---|---|---|
| 0 | air | 0.00 | 0.00 | 0.00 | 0.0 |
| 1 | tail | 2.04 ± 0.00 | 8.2 | 0.60 ± 0.00 | 0.0 |
| 2 | head $j$ = 0.5 | 0.68 | 14.98 | 0.41 ± 0.00 | 0.12 |
| 3 | $H_2O$ | 0.66 ± 0.00 | 9.41 | 0.41 ± 0.00 | 0.0 |
| 4 | head $j$ = 0.5 | 0.68 | 14.98 | 0.41 ± 0.00 | 0.12 |
| 5 | tail | 1.47 ± 0.00 | 8.2 | 0.41 ± 0.00 | 0.0 |
| 6 | tail $CH_3$ | 0.61 ± 0.01 | 6.2 | 0.16 ± 0.01 | 0.0 |
| 7 | tail | 1.86 ± 0.00 | 8.2 | 0.16 ± 0.01 | 0.0 |
| 8 | head $j$ = 0 | 0.76 | 14.45 | 0.41 ± 0.00 | 0.24 |
| 9 | $H_2O$ | 0.16 ± 0.00 | 9.41 | 0.41 ± 0.00 | 0.0 |
| 10 | $SiO_2$ | 1.76 ± 0.00 | 18.82 | 0.41 ± 0.00 | 0.0 |
| 11 | Si | 0.00 | 19.35 | 0.01 | 0.0 |

### 2.2.5 Sample 11: DSPC-DPPS50-DPPS50 ($j = 0.5$)-CaCl$_2$ 0.05 mM

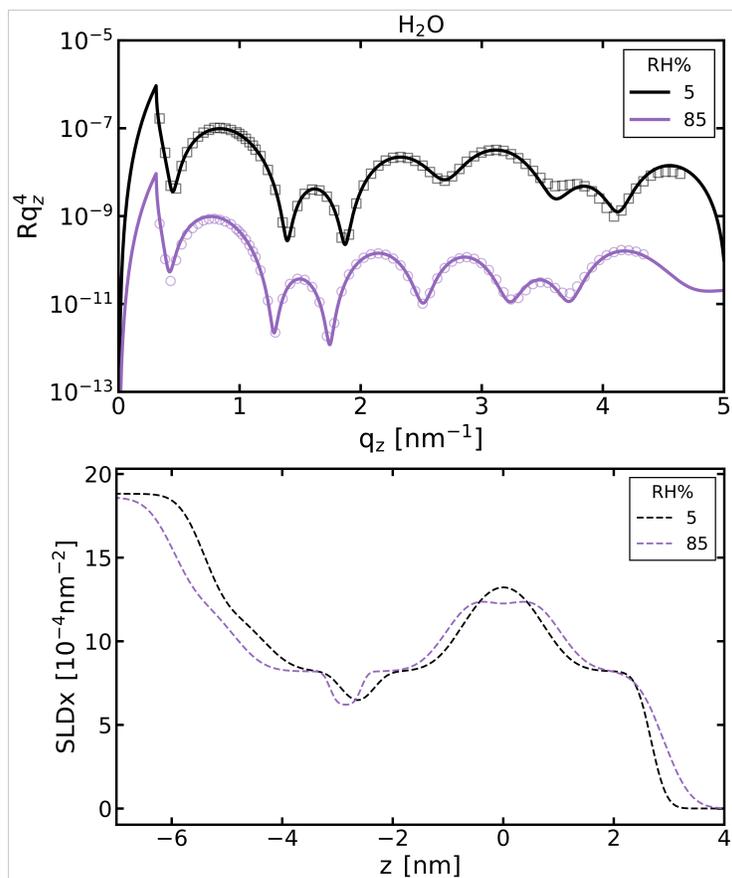

Figure S12: (Top) XRR data of a DSPC-DPPS50-DPPS50 trilayer in $H_2O$-CaCl$_2$ 0.05 mM solution ((Sample 11)) for 2 different relative humidity. Data are presented as $Rq_z^4$ on a semilog scale. Solid lines represent the best fits. (Bottom) SLD$_x$ profiles corresponding to the fits from above.

Table S49: Sample 11, parameters at RH = 5%

| # | Slab | Thickness (nm) | SLD ($\times 10^{-4}$ nm$^{-2}$) | Roughness (nm) | vfsolv (%) |
|---|---|---|---|---|---|
| 0 | air | 0.00 | 0.00 | 0.00 | 0.0 |
| 1 | tail | 1.95 ± 0.00 | 8.2 | 0.22 ± 0.00 | 0.0 |
| 2 | head $j = 0.5$ | 0.73 | 15.0 | 0.44 ± 0.00 | 0.22 |
| 3 | $H_2O$ | 0.00 ± 0.00 | 9.41 | 0.44 ± 0.00 | 0.0 |
| 4 | head $j = 0.5$ | 0.73 | 15.0 | 0.44 ± 0.00 | 0.22 |
| 5 | tail | 1.61 ± 0.00 | 8.2 | 0.44 ± 0.00 | 0.0 |
| 6 | tail $CH_3$ | 0.60 ± 0.01 | 6.2 | 0.20 ± 0.00 | 0.0 |
| 7 | tail | 1.47 ± 0.00 | 8.2 | 0.20 ± 0.00 | 0.0 |
| 8 | head $j = 0$ | 0.62 | 14.5 | 0.44 ± 0.00 | 0.35 |
| 9 | $H_2O$ | 0.34 ± 0.00 | 9.41 | 0.44 ± 0.00 | 0.0 |
| 10 | $SiO_2$ | 1.76 ± 0.01 | 18.82 | 0.38 ± 0.00 | 0.0 |
| 11 | Si | 0.00 | 19.35 | 0.01 | 0.0 |

Table S50: Sample 11, parameters at RH = 85%

| # | Slab | Thickness (nm) | SLD ($\times 10^{-4}$ nm$^{-2}$) | Roughness (nm) | vfsolv (%) |
|---|---|---|---|---|---|
| 0 | Air | 0.00 | 0.00 | 0.00 | 0.0 |
| 1 | tail | 1.96 ± 0.00 | 8.2 | 0.38 ± 0.00 | 0.0 |
| 2 | head $j = 0.5$ | 0.73 | 14.98 | 0.44 ± 0.00 | 0.14 |
| 3 | $H_2O$ | 0.42 ± 0.00 | 9.41 | 0.44 ± 0.00 | 0.0 |
| 4 | head $j = 0.5$ | 0.73 | 14.98 | 0.44 ± 0.00 | 0.14 |
| 5 | tail | 1.63 ± 0.01 | 8.2 | 0.44 ± 0.00 | 0.0 |
| 6 | tail $CH_3$ | 0.56 ± 0.01 | 6.2 | 0.10 ± 0.02 | 0.0 |
| 7 | tail | 1.78 ± 0.01 | 8.2 | 0.10 ± 0.02 | 0.0 |
| 8 | head $j = 0$ | 0.72 | 14.45 | 0.44 ± 0.00 | 0.29 |
| 9 | $H_2O$ | 0.27 ± 0.01 | 9.41 | 0.44 ± 0.00 | 0.0 |
| 10 | $SiO_2$ | 1.94 ± 0.01 | 18.6 ± 0.00 | 0.41 ± 0.00 | 0.0 |
| 11 | Si | 0.00 | 19.35 | 0.01 | 0.0 |

### 2.2.6 Sample 12: DSPC-DPPS-DPPS ($j = 1$)-CaCl$_2$

0.1 mM

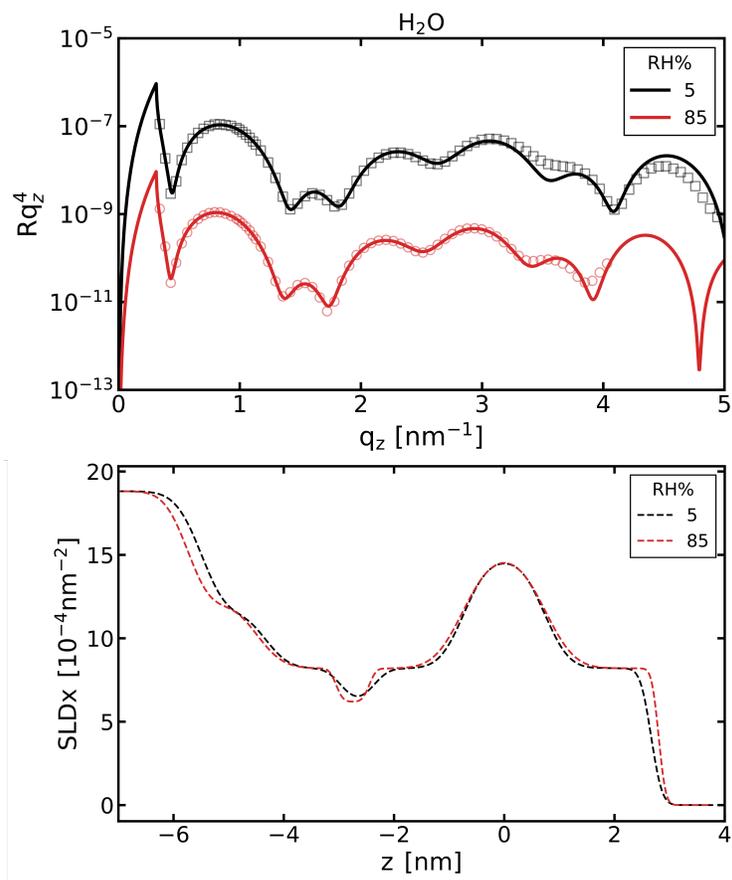

Figure S13: (Top) XRR data of a DSPC-DPPS-DPPS trilayer in H$_2$O-CaCl$_2$ 0.1 mM solution (Sample 12) for 2 different relative humidity. Data are presented as $\tilde{R}q_z^4$ on a semilog scale. Solid lines represent the best fits. (Bottom) SLD$_x$ profiles corresponding to the fits from above.

Table S51: Sample 12, parameters at RH = 5%

| # | Slab | Thickness (nm) | SLD ($\times 10^{-4}$ nm$^{-2}$) | Roughness (nm) | vfsolv (%) |
|---|---|---|---|---|---|
| 0 | air | 0.00 | 0.00 | 0.00 | 0.0 |
| 1 | tail | 1.93 ± 0.00 | 8.2 | 0.16 ± 0.00 | 0.0 |
| 2 | head $j = 1$ | 0.74 | 15.5 | 0.35 ± 0.00 | 0.13 |
| 3 | H$_2$O | 0.00 ± 0.00 | 9.41 | 0.35 ± 0.00 | 0.0 |
| 4 | head $j = 1$ | 0.74 | 15.5 | 0.35 ± 0.00 | 0.13 |
| 5 | tail | 1.63 ± 0.00 | 8.2 | 0.35 ± 0.00 | 0.0 |
| 6 | tail CH3 | 0.60 ± 0.01 | 6.2 | 0.22 ± 0.00 | 0.0 |
| 7 | tail | 1.40 ± 0.00 | 8.2 | 0.22 ± 0.00 | 0.0 |
| 8 | head $j = 0$ | 0.59 | 14.5 | 0.35 ± 0.00 | 0.47 |
| 9 | H$_2$O | 0.46 ± 0.00 | 9.41 | 0.35 ± 0.00 | 0.0 |
| 10 | SiO$_2$ | 1.92 ± 0.00 | 18.82 | 0.41 ± 0.00 | 0.0 |
| 11 | Si | 0.00 | 19.35 | 0.01 | 0.0 |

Table S52: Sample 12, parameters at RH = 85%

| # | Slab | Thickness (nm) | SLD (×$10^{-4}$ nm$^{-2}$) | Roughness (nm) | vfsolv (%) |
|---|---|---|---|---|---|
| 0 | air | 0.00 | 0.00 | 0.00 | 0.0 |
| 1 | tail | 1.96 ± 0.00 | 8.2 | 0.38 ± 0.00 | 0.0 |
| 2 | head $j = 1$ | 0.73 | 14.98 | 0.44 ± 0.00 | 0.14 |
| 3 | H$_2$O | 0.42 ± 0.00 | 9.41 | 0.44 ± 0.00 | 0.0 |
| 4 | head $j = 1$ | 0.73 | 14.98 | 0.44 ± 0.00 | 0.14 |
| 5 | tail | 1.63 ± 0.01 | 8.2 | 0.44 ± 0.00 | 0.0 |
| 6 | tail CH$_3$ | 0.56 ± 0.01 | 6.2 | 0.10 ± 0.02 | 0.0 |
| 7 | tail | 1.78 ± 0.01 | 8.2 | 0.10 ± 0.02 | 0.0 |
| 8 | head $j = 0$ | 0.72 | 14.45 | 0.44 ± 0.00 | 0.29 |
| 9 | H$_2$O | 0.27 ± 0.01 | 9.41 | 0.44 ± 0.00 | 0.0 |
| 10 | SiO$_2$ | 1.94 ± 0.01 | 18.6 | 0.41 ± 0.00 | 0.0 |
| 11 | Si | 0.00 | 19.35 | 0.01 | 0.0 |

## 2.3 Area per molecule

From the fitted tail thickness, we estimated the area per molecule (APM) or area per lipid. It is calculated as APM = $V_m/t$, where $V_m$ = 825 Å$^3$ is the molecular volume of the DP tail and ( $t$ is the tail thickness. Figure S14 shows APM values for all samples studied by reflectivity. The APM values were all close to 42 Å$^2$ and were consistent with the molecular area at the air/water interface prior to transfer and the numerical simulations (see below).

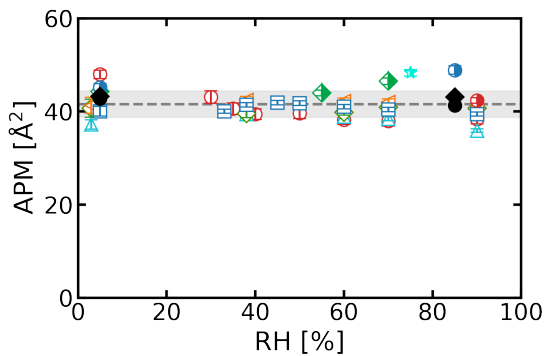

Figure S14: Area per molecule values for all samples studied by reflectivity. All markers follow the nomenclature used in the main manuscript. The mean value of 41.6 Å$^2$ is represented by a grey dashed line. The grey band corresponds to $\pm\sigma$ = 2.8 Å$^2$.

# 3 Simulations Details

## 3.1 Model

All simulations are performed using the Gromacs package versions 2023.4 or 2024.3 **?**, using the C36-CHARMM36 force field **?** for 1,2-dipalmitoyl-sn-glycero-3-phosphoserine (DPPS, negatively charged) and sodium (Na$^+$), with the OPC water model **?**. The C36-CHARMM force field was shown to describe at least partially some of the experimental features of the serine headgroup flexibility **?**, and the specificity of interfacial water dynamics nearby lipidic bilayers **?**. The CHARMM family of force fields were developed for use with the mTIP3P water model **?**, which has an anomalously high dielectric constant **?**. It has been shown that the CHARMM36 lipids could also be used with more precise water models (TIP4P and OPC) **??**. The water was therefore converted to the OPC model. Each system is characterized by its Hydration Number (HN): the number of water molecule per lipid, that varied from 0 to 45. For all systems, 3D-periodic boundary conditions and a time step of 2 fs are employed. Lennard–Jones interactions are truncated at $r_c$ = 1.2 nm. Starting from a switching distance of 1.0 nm, the force was switched smoothly to 0 at the cutoff distance of 1.2 nm as implemented in the Gromacs Software. For the electrostatic interactions the Particle Mesh Ewald method is employed with a real-space cut-off $r_c$ = 1.2 nm **?**. The Verlet list is updated using Gromacs' default energy tolerance of 0.005kJ/mol/ps. Three thermostats were always used, one for each monolayer, and one for the solvent group (water+sodium). A leapfrog integrator was used with a 2 fs time step. Water bonds were constrained to their equilibrium values using the LINCS algorithm **?**. Trajectory frames were stored at 10 ps intervals.

*Initial Systems Construction.* To mimic the experimental setup, the system was first constructed as a gel phase monolayer, and converted into a bilayer by assembling two monolayers (see Fig. S15). The

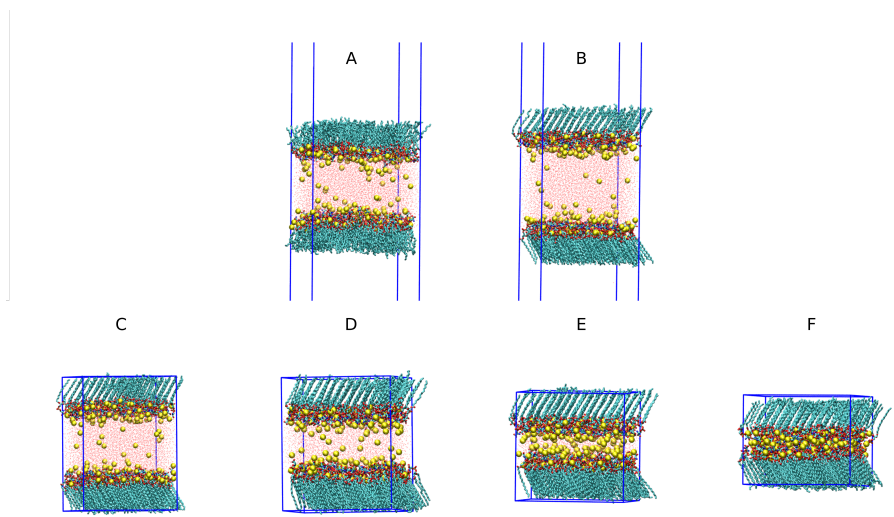

Figure S15: Typical snapshots of the simulated systems. Periodic boundary conditions are used in the three dimensions, with simulation boxes depicted by blue frames. A : Two monolayers at the solvent/vacuum interface in the initial conformation of a $L_\alpha$ fluid phase at high hydration level, constructed from a bilayer pre-equilibrated by CHARMM-GUI **?**, transformed into two monolayers by increasing the box in one direction ($L_z$ = 30 nm). B : Same system as A, after cooling for 2 μs towards 293K, the lipids are in a $L'_\beta$ gel phase. C, D, E, F : final bilayer systems after dehydration, equilibration and production, for hydration numbers of 35, 25, 15 and 5 respectively. Water molecules are represented by red dots at the position of the oxygen, Sodium ions by big yellow spheres, and the bonds of DPPS lipids by cylinders. Hydrogens are not shown. Illustration created using VMD **?**.

first system contained two monolayers of 100 DPPS each, disposed on the upper and lower interface of layer of solvent composed of 9000 water molecules and 200 sodium ions. This initial system was generated using the CHARMM-GUI **?**. This system was pre-equilibrated in the fluid phase in a $NL_zP_{xy}T$ semi-isotropic ensemble with $P_{xy}$ = 1 bar, $L_z$ = 30 nm and $T$ = 333 K (Fig. S15A). It was then cooled from 333 K to 293 K with a cooling rate of 20 K per μs (Nose-Hoover thermostat, Parrinello-Rahman barostat with a time constant of 5 ps and compressibility of $4.5 \times 10^{-5}$ per bar). The resulting conformation represents a $L_\beta$ (gel) phase at a temperature of $T$ = 293 K, with a hydration level of $HN$ = 45 and an area per lipid of 44.21 Å$^2$ (Fig. S15B). The monolayers were gradually dehydrated by removing 1, 3 or 5 water molecules per lipid at a time. Each dehydration was followed by an equilibration of 10 ns ($NP_zA_{xy}T$ ensemble with $P_z$ = 1 bar and $T$ = 293 K using Nose-Hoover thermostat and Berendsen barostat with a time constant of 5 ps) before a new dehydration was performed. Finally, the vacuum between the monolayer was removed by diminishing by hand the box length $L_z$ to generate the bilayers with varying hydration numbers.

*Equilibration and Production.* Each of the bilayer systems was equilibrated during 1000 ns in an $NP_zA_{xy}T$ ensemble with $P_z$ = 1 bar and $T$ = 293 K using Nose-Hoover thermostat and Berendsen barostat with a time constant of 5 ps. The area per lipid is kept constant. The production is then

performed for 600 ns in the $NL_zA_{xy}T$ ensemble with $T = 293$ K using Nose-Hoover thermostat and no barostat (see typical systems in Fig. S15C-F).

*Analysis.* The SLDn and SLDx profiles in the direction normal to the bilayers presented in Figure 2C were obtained by summing the products of the number density of each atom by its corresponding neutron or X-ray scattering. These calculations were performed using the Gromacs `density` tool with a fixed bin size of 0.1 Å, after centering on the center of mass of the solvent group (water and ions) at each timestep. When computing the SLDn profiles with $D_2O$, we considered that all $H_2O$ molecules had been replaced with $D_2O$. For each hydration number, 60 000 configurations were analysed (one every 10 ps). The polarization profiles $P_z(z)$ in the direction normal to the bilayers presented in Figure S16 were calculated with a home-made script using the MDAnalysis package **?**. For each hydration number, 6000 configurations were analysed (one every 100 ps). The error bars were obtained from the standard error of the mean over 6 analyses of blocks of 100 ns each. The vectorial sum of the dipole moments of the OPC molecules was calculated with a dipole 0.0516 e.nm per molecule (2.48 D). For the comparison with the Landau-Ginzburg (LG) model proposed by Schlaich et al.**?**, the region for the fit was first selected on the basis of the local density of OPC molecules ($\rho_{OPC}(z) \geq 32$ molecule/nm$^3$), and the thickness of this region, was used for the $d$ parameter of the LG model. Within this region, the profiles were fitted with two different approaches. In a first step, each of the curve was fitted using least-square minimization (`curve_fit` in scipy **?**) with the equation

$$m_z(z, d) = m_0 \frac{\sinh(z/\lambda_H)}{\sinh(d/(2\lambda_H))}, \tag{S2}$$

with two parameters ($m_0$, $\lambda_H$) per curve, ie per $d$ value.

The resulting fits and the resulting parameters as a function of $d$ are illustrated in Fig. S16. For *HN* lower than 15, the parameters are not given, since the curves are so close to a straight line that $\lambda_H$ cannot be defined accurately using a least squares fit. These curves were considered as non-informative. In a second step, the curves with $HN \geq 20$ were all fitted together with the equation :

$$m_z(z, d) = (-h/a) \frac{1}{2\lambda_H} \frac{\sinh(z/\lambda_H)}{\cosh(d/(2\lambda_H))}, \tag{S3}$$

with only two parameters ($-h/a$, $\lambda_H$) for all curves. The resulting fits are depicted in Fig. S16-A, with the optimized parameters $-h/a = 0.46$ e.nm$^{-1}$ and $\lambda_H = 0.48$ nm in Fig. S16-B and C.

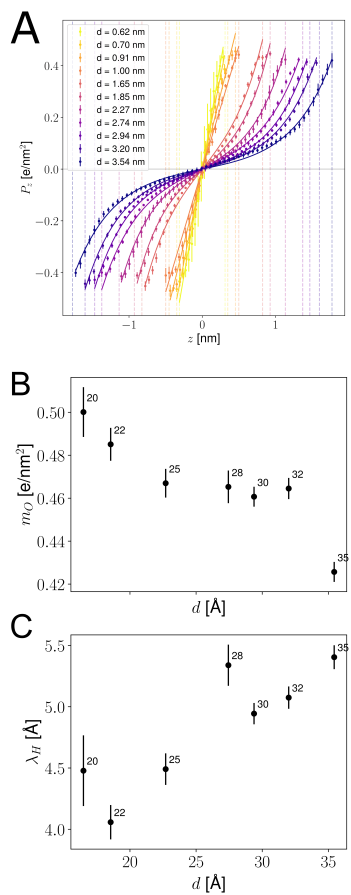

Figure S16: (A) Water dipolar polarization profiles, within the water layer of thickness *d*, for systems of different hydration numbers. The dashed vertical lines indicate the surface positions at $\pm d/2$, (*d* are given in the legends). The rest of the profile, not used for the fitting, is not represented. Solid lines are fits according to Eq. S2 leading to the parameters $m_0$ and $\lambda_H$. (B) and (C) Variations of the parameters $m_0$, $\lambda_H$ for different hydration level, ie different *d*. The points are labelled with the number of water per lipid. The error bars are obtained from the square root of the covariances of the fit.

# 4 X-ray Fluorescence Measurements

The number and the nature of the counterions attached to the DPPS lipid were determined using Total Reflection X-ray Fluorescence (TXRF). A single-element Silicon Drift Detector (XFlash 430M, Bruker, Germany) was mounted directly above the sample at a 90° angle. The detector was separated from the helium chamber by a 13 $\mu$m Kapton window and a 1 cm air gap. The X-ray beam has an energy of 8 keV and a glancing angle of 0.18°, below the critical angle for total reflection (0.22°). This ensured that only an evanescent wave penetrated the Si wafer, thereby reducing background noise and minimizing the contribution of the Si peak.

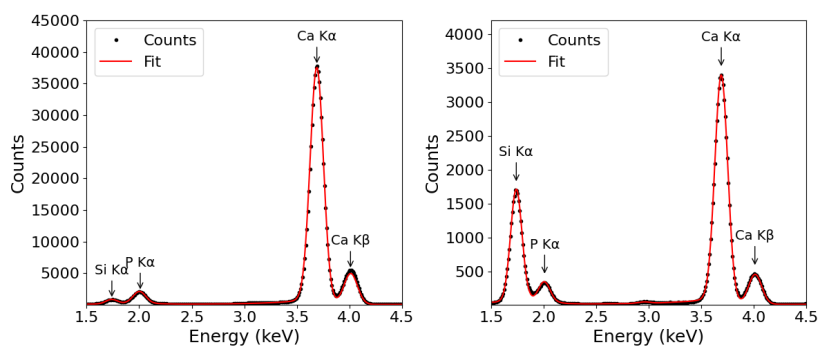

Figure S17: Left: X-ray fluorescence spectrum measured on the dried thin film of $Ca(H_2PO_4)_2$ used for calibration. Right: X-ray fluorescence spectrum measured on trilayer DSPC-2 DPPS samples with 80% DPPS and 20% DPPC in the second and third layers. The red lines represent the best fits obtained using PyMca.

In the TXRF configuration, the area of each elemental peak is proportional to the quantity of the corresponding element. To determine the $Ca^{2+}$ content, calibration was performed using a 0.8 mM solution of $Ca(H_2PO_4)_2$ in water. A few microliters of this solution were deposited onto a clean Si substrate and allowed to dry in a helium atmosphere. XRF spectra were fitted using the PyMca software ? for deconvolution of the elemental peaks. This calibration step enabled the determination of the ratio between the peak areas associated with P and Ca for a solution containing a 2:1 P-to-Ca ratio. This ratio was then used to quantify P-to-Ca ratios in our samples without additional assumptions (see Fig. S17).

We applied this analysis to samples prepared with a 0.05 mM $Ca(OH)_2$ subphase, using different DPPS-to-DPPC ratios in the second and third monolayers: 100% DPPS, 80% DPPS, and 50% DPPS (see Fig. S18). The number of Ca ions per DPPS lipid remained close to 0.5, regardless of the DPPS-to-DPPC ratio. This value matches the expected ratio of one divalent $Ca^{2+}$ ion for every two PS head groups.

It is known that charged monolayers at the air-water interface can attract calcium ions present as traces in ultrapure water ?. Using the same approach, we quantified the amount of Ca per P in samples prepared in ultrapure water without added salt, obtaining a maximum value of 0.17 Ca per P. Assuming that all Ca ions are present as $Ca^{2+}$ bound to PS head groups, this corresponds to a maximum value of 0.25 Ca per DPPS.

Finally, we verified that DPPS counterions could be replaced by specific ions introduced into the subphase before deposition. We prepared trilayer samples on a subphase containing 0.1 mM CsOH, using two DPPS-to-DPPC ratios (1 and 0.5). Although we lack a reference for absolute calibration in this case, the TXRF spectra clearly showed a strong Cs signal in the sample (see Fig. S19). The ratio of the Cs peak areas between 50% DPPS and 100% DPPS was 0.54, strongly suggesting that Cs ions bind to the PS head groups in a one-to-one ratio. Furthermore, the amount of Ca was significantly lower in these

samples, with a maximum of only 0.003 Ca-per-DPPS detected.

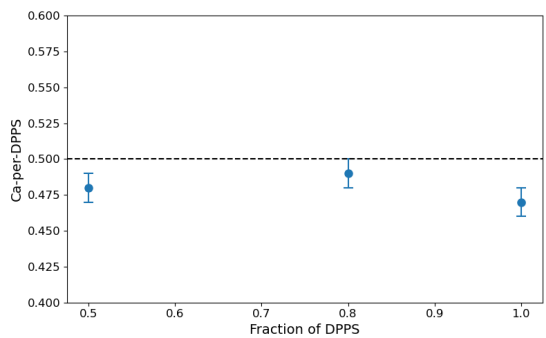

Figure S18: Evolution the number of Ca atoms per DPPS lipid measured by X-ray fluorescence. The line represents the theoretical value of 1 $Ca^{2+}$ for 2 PS heads.

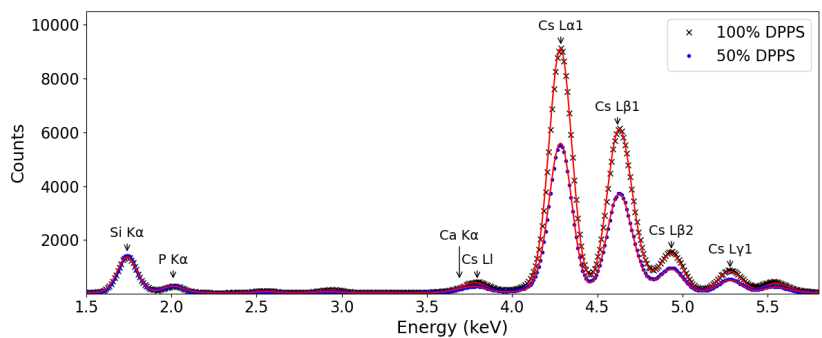

Figure S19: X-ray fluorescence spectra measured on trilayer DSPC-2 DPPS samples prepared on 1 mmol/L CsOH, with 100% DPPS (gray crosses) and 50% DPPS (blue circles) in the second and third layers. The red and magenta lines represent the best fits obtained using PyMca.

71

## 5 Charge positions and $z_{cc}$ determination

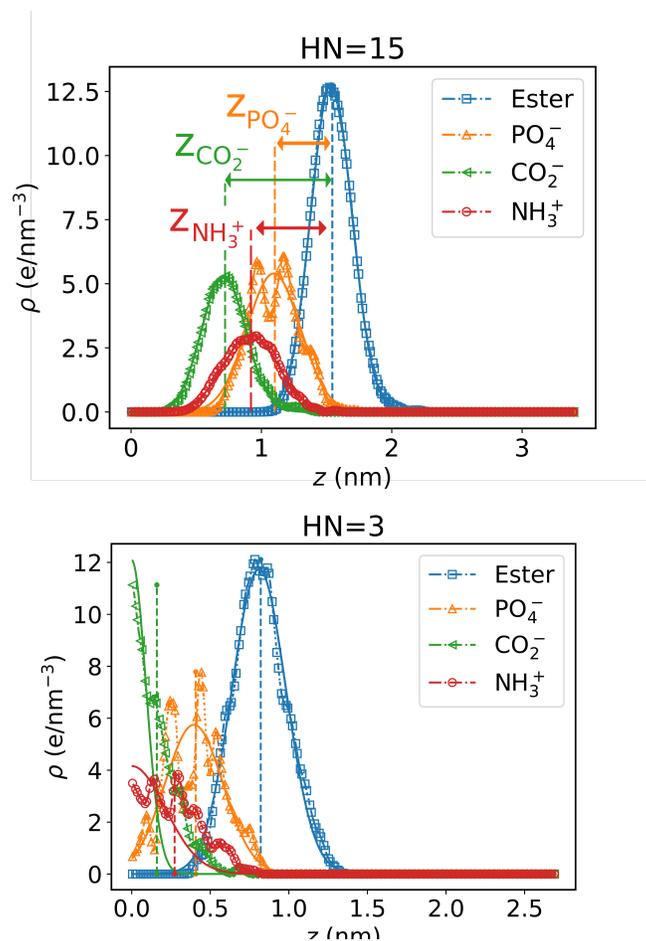

Figure S20: Absolute value of the charge density from numerical simulations, and best fits by Gaussian functions, for the oxygens of the ester group (charge −2.24$e$), phosphate group $PO_4^-$ (charge −1.2$e$), carboxylic group $COO^-$ (charge −1.0$e$) and ammonium group $NH_3^+$ (charge +0.7$e$). (Top) Hydrated limit (HN=15). (Bottom) Highly dehydrated limit (HN=3). On the top graph the position of the different charged chemical groups are defined.

In this section, we justify the definition of the offset $z_{sh}$ used to calculate the distance between the charged planes $z_{cc}$. From numerical simulations, it is possible to extract the charge density of the various chemical groups. Here, we focus mainly on four groups present in serine heads: the ester group (represented by their oxygens, charged −2.24$e$), the phosphate group (negatively charged −$e$), the

72

carboxyl group (negatively charged $-e$) and the ammonium group (positively charged $+e$). The head has the overall charge $-e$.

We have plotted on figure S20 the charge densities corresponding to these four groups, averaged laterally over the size of the simulation box, for two hydration numbers (HN=3 and HN=45). By performing a Gaussian fit of these distributions, it is possible to obtain the average position of the charges associated with these groupings. We have chosen to measure these positions relative to the position of the ester group, which marks the transition between the chains and the lipid heads, and corresponds to the zone of maximum change in the contrast (both for $SLD_X$ and $SLD_n$). These positions are noted $z_{PO_4^-}$, $z_{CO_2^-}$ and $z_{NH_3^+}$ (see Figure S20).

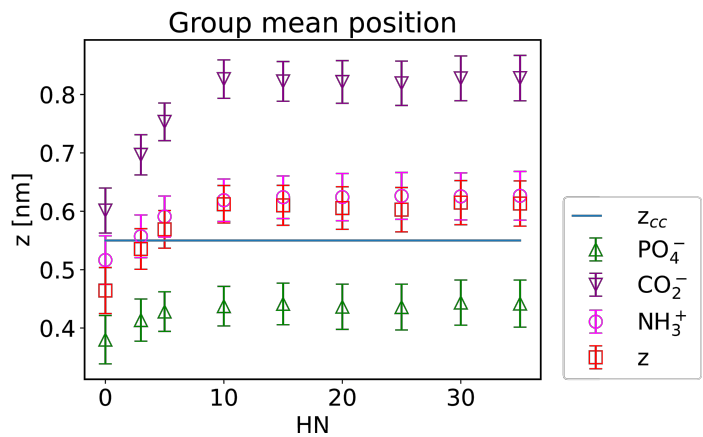

Figure S21: Average positions of the various charged groups $z_{PO_4^-}$, $z_{CO_2^-}$ and $z_{NH_3^+}$. Red square corresponds to the position of the barycentre of the charge with a ponderation corresponding to the partial charges of each groups: -1.2 ($PO_4^-$), -1.0 ($CO_2^-$) and +0.7 ($NH_3^+$).

The results obtained for the average positions of the various charged groups are shown in Figure S21 for different hydration numbers. Firstly, it can be seen that these positions depend very little on the hydration of the system, although for low hydrations the positions of the $CO_2^-$ and $NH_3^+$ groups, the most external to the lipid layer, are less well resolved as the two monolayers are extremely close.

We can see that the barycentre of the two negative groups is very close to the position of the positive charge. We've also plotted the position of the barycentre of the charge set (red squares), which carries a negative charge $-e$. This leads us to define a $z_{sh} = 0.55 \pm 0.1$ nm, which enables us to calculate the distance between the charged planes $z_{cc}$.

## 6 Interaction potentials and strong coupling theory

### 6.1 Interaction potentials

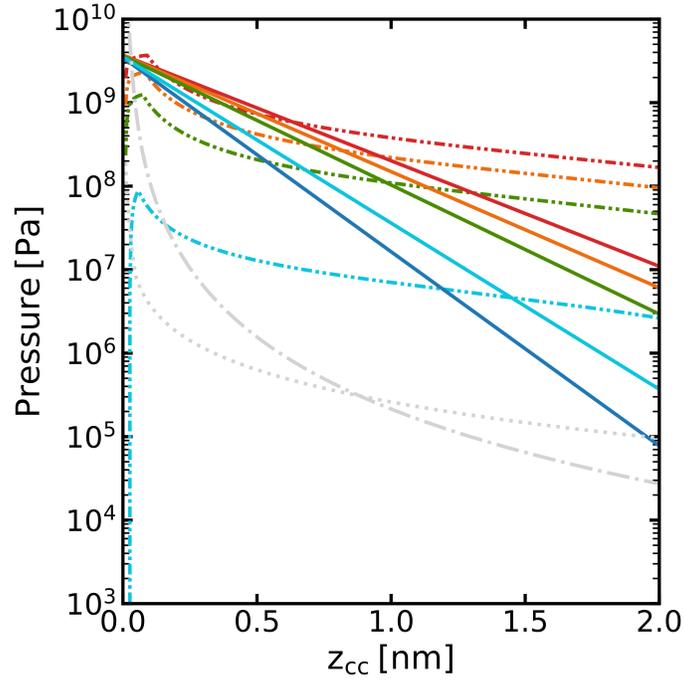

Figure S22: Representation of the different contributions to pressure for different charge fractions. For attractive pressure as $P_{vdW}$ and $P_{SC}$, the absolute value is taken. (Solid lines) hydration pressure (equation 7 in main paper); $(-\cdot\cdot-)$ strong coupling (equation 6 in main paper); $(-\cdot-)$ van der Waals $P_{vdW} = -H/6\pi d^3$. The Hamaker constant is $H \sim 2 \cdot 10^{-21}$ J; $(\cdots)$ $P_{SC}$ weak coupling electrostatic interactions. Electrostatic pressures are calculated using $\epsilon_r$ calculated with 5 in main paper. $P_{WC}$ is computed using the surface charge density $\sigma_S$, corresponding to $j = 0.2$.

The different contributions to the interaction potential between charged surfaces are shown on Figure S22 for different charge densities. Below 1 nm, van der Waals and weak coupling interactions are always one or two order of magnitudes smaller than hydration and strong coupling interactions.


  
\end{document}